\documentclass[11pt]{article}

\makeatletter\renewcommand{\section}{\@startsection
{section}{1}{\z@}{-3.5ex plus -1ex minus
    -.2ex}{2.3ex plus .2ex}{\large\bf }}

\makeatletter\renewcommand{\subsection}{\@startsection{subsection}{2}{\z@}{-3.25ex
plus -1ex minus
   -.2ex}{1.5ex plus .2ex}{\bf }}
\makeatletter\renewcommand{\subsubsection}{\@startsection{subsubsection}{3}{-2.45ex}{-3.25ex
plus -1ex minus -.2ex}{1.5ex plus .2ex}{\it }}

\renewcommand*\l@section{\@dottedtocline{1}{0em}{1.5em}}
\renewcommand*\l@subsubsection{\@dottedtocline{4}{3.8em}{3.2em}}

\renewcommand\tableofcontents{%
    \section*{\Large\contentsname
        \@mkboth{%
           \MakeUppercase\contentsname}{\MakeUppercase\contentsname}}%
       {\baselineskip=15pt plus 2pt minus 1pt
    \@starttoc{toc}}%
\vspace{-3mm}\centerline{{\vrule height 0.5pt width 15.5cm depth
0pt}} }

\renewenvironment{thebibliography}[1]
     {
      \section*{{\refname}
        \@mkboth{\MakeUppercase\refname}{\MakeUppercase\refname}}%
     \list{\@biblabel{\@arabic\c@enumiv}}%
           {\settowidth\labelwidth{\@biblabel{#1}}%
            \leftmargin\labelwidth
            \advance\leftmargin\labelsep
            \@openbib@code
            \usecounter{enumiv}%
            \let\p@enumiv\@empty
            \renewcommand\theenumiv{\@arabic\c@enumiv}}%
      \sloppy
      \clubpenalty4000
      \@clubpenalty \clubpenalty
      \widowpenalty4000%
      \sfcode`\.\@m}

\usepackage{amsfonts}
\usepackage{amssymb}
\usepackage{mathrsfs}
\usepackage{amsmath,amssymb}
\usepackage{bbm}

\numberwithin{equation}{section}

\newcommand{\IN}{\mathbbm{N}} 
\newcommand{\IR}{\mathbbm{R}}    
\newcommand{\IC}{\mathbbm{C}}    
\newcommand{\IZ}{\mathbbm{Z}}    

\newcommand{\CA}{\mathcal{A}}

\newcommand{\CC}{\mathcal{C}}
\newcommand{\CE}{\mathcal{E}}
\newcommand{\CF}{\mathcal{F}}    
\newcommand{\CL}{\mathcal{L}}    
    
\newcommand{\CN}{\mathcal{N}} 
\newcommand{\CO}{\mathcal{O}} 
\newcommand{\CP}{\mathcal{P}} 
\newcommand{\CT}{\mathcal{T}} 
\newcommand{\CU}{\mathcal{U}} 
\newcommand{\CW}{\mathcal{W}} 

\newcommand{\hCW}{\hat{\CW}} 

\renewcommand{\a}{\alpha}
\newcommand{\ad}{{\dot\a}}
\renewcommand{\b}{\beta}
\newcommand{\bd}{{\dot\b}}
\newcommand{\g}{\gamma}

\renewcommand{\d}{\delta}

\newcommand{\te}{\theta}   
\newcommand{\tx}{{\tilde{x}}}
\newcommand{\dt}{{\rm d}}

\newcommand{\der}[1]{\frac{\partial}{\partial #1}}

\begin{document}
\begin{titlepage}
\setcounter{page}{0}
\begin{flushright}
  hep-th/0608225\\
  ITP--UH--24/06
\end{flushright}
\vskip 2.0cm
\begin{center}
{\LARGE \bf Hidden Symmetries and Integrable Hierarchy\\[8pt]
of the $\CN=4$ Supersymmetric Yang-Mills Equations}\\
\vskip 2.0cm
\renewcommand{\thefootnote}{\fnsymbol{footnote}}
{\Large Alexander D. Popov\footnote{On leave from Bogoliubov
Laboratory of Theoretical Physics, JINR, Dubna, Russia.}
and Martin Wolf\footnote{Address after October 1st, 2006:
Theoretical Physics Group,
The Blackett Laboratory, Imperial College London, Prince
Consort Road, London SW7 2BW, United Kingdom.}
} \setcounter{footnote}{0}
\renewcommand{\thefootnote}{\arabic{thefootnote}}
\vspace{.8cm}

{\em Institut f\"ur Theoretische Physik\\
     Leibniz Universit\"at Hannover\\
     Appelstra{\ss}e 2, 30167 Hannover, Germany}
\vspace{.8cm}

 {E-mail: {\ttfamily popov, wolf@itp.uni-hannover.de} } \vspace{0.8cm}
\end{center}
\begin{center}
{\bf Abstract}
\end{center}
\begin{quote}

We describe an infinite-dimensional algebra of hidden 
symmetries of $\CN=4$ supersymmetric Yang-Mills (SYM) theory. 
Our derivation is based on a generalization of the 
supertwistor correspondence. Using the latter, we 
construct an infinite sequence of flows on the 
solution space of the $\CN=4$ SYM equations. The dependence 
of the SYM fields on the parameters 
along the flows can be 
recovered by solving the equations of the hierarchy.
We embed the $\CN=4$ SYM equations in the infinite system 
of the hierarchy equations and show that this SYM hierarchy 
is associated with an infinite set of graded symmetries 
recursively generated from supertranslations. Presumably, 
the existence of such nonlocal symmetries underlies
the observed integrable structures in quantum $\CN=4$ 
SYM theory.

\end{quote}
\end{titlepage}

\section{Introduction}

Let us consider an open subset $\CP^3:=\IC P^3\setminus\IC P^1$ 
of the three-dimensional complex projective space $\IC P^3$ 
together with the following double fibration:
\begin{equation}\label{eq:DF1}
\begin{aligned}
\begin{picture}(50,40)
\put(0.0,0.0){\makebox(0,0)[c]{$\CP^3$}}
\put(64.0,0.0){\makebox(0,0)[c]{$\IC^4$}}
\put(34.0,33.0){\makebox(0,0)[c]{$\IC^4\times\IC P^1$}}
\put(7.0,18.0){\makebox(0,0)[c]{$$}}
\put(55.0,18.0){\makebox(0,0)[c]{$$}}
\put(25.0,25.0){\vector(-1,-1){18}}
\put(37.0,25.0){\vector(1,-1){18}}
\end{picture}
\end{aligned}
\end{equation}
Here, $\IC^4$ is complexified Minkowski space and $\CP^3$ its
associated {\it twistor space} \cite{Penrose:1969}. It can be 
shown that $\CP^3$ is a rank $2$ holomorphic vector bundle 
$\CP^3=\CO(1)\oplus\CO(1)$ over the Riemann sphere $\IC P^1$, 
where $\CO(n)$ denotes the holomorphic line bundle 
$\CO(n)\to\IC P^1$ parametrized by the first Chern number 
$n\in\IZ$. The importance of the diagram \eqref{eq:DF1} lies
 in the fact that there is a correspondence between 
holomorphic vector bundles over $\CP^3$ obeying certain
triviality conditions and
holomorphic vector bundles over $\IC^4$ equipped with a 
connection satisfying the self-dual Yang-Mills (SDYM) 
equations on $\IC^4$ \cite{Ward:1977ta}. This map between
vector bundles
has been termed the {\it Penrose-Ward transform}
\cite{MasonRF}.

Substituting the twistor space $\CP^3$ by 
$\CP^3_n:=\CO(n)\oplus\CO(n)$ (for $n> 1$)
which has the same dimension but different topology, one 
obtains the double fibration
\begin{equation}\label{eq:DF2}
\begin{aligned}
\begin{picture}(50,40)
\put(0.0,0.0){\makebox(0,0)[c]{$\CP^3_n$}}
\put(64.0,0.0){\makebox(0,0)[c]{$\IC^{2(n+1)}$}}
\put(34.0,33.0){\makebox(0,0)[c]{$\IC^{2(n+1)}\times\IC P^1$}}
\put(7.0,18.0){\makebox(0,0)[c]{$$}}
\put(55.0,18.0){\makebox(0,0)[c]{$$}}
\put(25.0,25.0){\vector(-1,-1){18}}
\put(37.0,25.0){\vector(1,-1){18}}
\end{picture}
\end{aligned}
\end{equation}
and the above-mentioned Penrose-Ward transform is generalized 
to a correspondence between holomorphic vector bundles over 
$\CP^3_n$ and solutions to a system of equations (called the 
SDYM hierarchy truncated up to level $n$ \cite{MasonRF}) for 
Yang-Mills-Higgs fields on the space
\begin{equation}\label{eq:genspti}
 \IC^{2(n+1)}\ \cong\ \IC^4\times\IC^{2(n-1)}.
\end{equation}
Here, $\IC^4$ is interpreted as (complexified) space-time 
and $\IC^{2(n-1)}$ as a space of ``higher times" parametrizing 
solutions to the SDYM equations which in turn
appear as a subset of the 
truncated SDYM hierarchy \cite{MasonRF}--\cite{Mason:1994rs}. 
Letting $n$ tend to infinity, one obtains the 
full SDYM hierarchy which is associated with an affine 
extension of translation symmetries \cite{MasonRF},
\cite{Popov:1995qb}--\cite{IvanovaZT}.

The above construction can be generalized to $\CN$-extended 
supersymmetric SDYM theory \cite{Semikhatov:1982ig} by substituting 
$\CP^3_n$ by a generalized supertwistor space \cite{Wolf:2004hp},
\begin{equation}\label{eq:sutwsp}
 \CP^{3|\CN}_n\ :=\ 
   \CO(n)\otimes\IC^2\oplus\Pi\CO(n)\otimes\IC^\CN,
\end{equation}
associated with the superspace $\IC^{2(n+1)|2\CN}$ 
extending the space \eqref{eq:genspti}. The operator $\Pi$ 
inverts the Grassmann parity of the fibre coordinates. In a 
seminal recent paper \cite{Witten:2003nn}, Witten observed that 
the supertwistor space $\CP^{3|4}:=\CP^{3|4}_1$ is a Calabi-Yau 
supermanifold and showed that B-type open topological string 
theory with the 
supertwistor space as target space 
(twistor string theory\footnote{For other variants of 
twistor string models see \cite{Berkovits:2004hg}.}) 
is equivalent to holomorphic
Chern-Simons (hCS) theory\footnote{Holomorphic Chern-Simons theory 
describes
(inequivalent) holomorphic structures on a vector 
bundle over a given complex (super)manifold, which in the case at hand 
is $\CP^{3|4}$.} on the same space. This theory is in turn 
equivalent to $\CN=4$ SDYM theory in four dimensions. 
Generalizing an earlier construction \cite{Nair:1988bq}, 
it was also shown in \cite{Witten:2003nn} that it is possible to 
recover perturbative $\CN=4$ supersymmetric Yang-Mills (SYM) 
theory in terms of integrals 
over moduli spaces of algebraic curves (D-instantons) in the 
supertwistor space $\CP^{3|4}$. For an account of progress made 
in this area, see e.g. \cite{webpage,Cachazo:2005ga} 
and references therein. For other aspects of twistor string 
theories discussed lately see e.g. 
\cite{Neitzke:2004pf}--\cite{Saemann:2006tt}.

A twistor construction was developed not only for 
(supersymmetric) SDYM 
theory but also for full $\CN$-extended SYM theory 
\cite{Witten:1978xx}--\cite{Howe:1995md}
with $0\leq\CN\leq 4$ (for recent reviews 
see \cite{Popov:2004rb,Saemann:2006tt}). 
Recall that the twistor description of $\CN=4$ SYM theory is 
based on the $\CN=3$ superspace formulation of the latter 
\cite{Witten:1978xx}. For that one considers the supertwistor
space $\CP^{3|3}$ as in \eqref{eq:sutwsp} with $n=1$, 
$\CN=3$ and the dual supertwistor space $\CP^{3|3}_*=\IC 
P^{3|3}_*\setminus\IC P^{1|3}_*=\CO(1)\otimes\IC^2_*\oplus
\Pi\CO(1)\otimes\IC^3_*$ which is fibred over $\IC P^1_*$. 
Let $\pi_\ad$ with $\ad=\dot1,\dot2$ and $\rho_\a$ 
with $\a=1,2$ be homogeneous coordinates on $\IC P^1$ and 
$\IC P^1_*$, respectively. Then the {\it superambitwistor space}, 
denoted by $\CL^{5|6}$, is defined as a quadric surface 
\begin{equation}
 z^\a\rho_\a-w^\ad\pi_\ad+2\te^i\eta_i\ =\ 0
\end{equation}
in $\CP^{3|3}\times\CP^{3|3}_*$ \cite{Witten:1978xx}. Here, 
$(z^\a,\pi_\ad,\eta_i)$ and $(w^\ad,\rho_\a,\te^i)$ are 
homogeneous coordinates on $\CP^{3|3}$ and
$\CP^{3|3}_*$, respectively, and $\eta_i$, $\te^i$ are 
Grassmann odd variables with $i=1,2,3$. Interestingly, 
$\CL^{5|6}$ is a Calabi-Yau supermanifold and it is possible 
to formulate twistor string theory on it \cite{Witten:2003nn}.

Note that if pr$_1$ and pr$_2$ are the two projections from 
$\IC P^1\times\IC P^1_*$ onto $\IC P^1$ and $\IC P^1_*$, then 
the space $\CL^{5|6}$ can be defined as a holomorphic vector 
bundle over $\IC P^1\times\IC P^1_*$ in terms of a short 
exact sequence of vector bundles
\begin{equation}\label{eq:seq1}
 0\ \to\ \CL^{5|6}\ \to\ {\rm pr}_1^*\CP^{3|3}
    \oplus{\rm pr}_2^*\CP^{3|3}_*\ 
    \overset{\kappa}{\to}\ \CO(1,1)\ \to\ 0 
\end{equation}
generalizing the analogous sequence for the purely bosonic 
case \cite{Burns,LeBrun:1983}. The mapping $\kappa$ is given
by $\kappa(z^\a,\pi_\ad,\eta_i,w^\ad,\rho_\a,
\te^i):=z^\a\rho_\a-w^\ad\pi_\ad+2\te^i\eta_i$ and 
$\CO(1,1):={\rm pr}^*_1\CO(1)\otimes{\rm pr}^*_2\CO(1)$. The
corresponding long exact cohomology sequence gives 
$H^1(\IC P^1\times\IC P^1_*,\CL^{5|6})=0$ and
$H^0(\IC P^1\times\IC P^1_*,\CL^{5|6})=\IC^{4|12}$. Hence, 
by virtue of the Kodaira theorem \cite{Kodaira:1962} we 
obtain the double fibration
\begin{equation}\label{eq:DF3}
\begin{aligned}
\begin{picture}(50,40)
\put(0.0,0.0){\makebox(0,0)[c]{$\CL^{5|6}$}}
\put(64.0,0.0){\makebox(0,0)[c]{$\IC^{4|12}$}}
\put(34.0,33.0){\makebox(0,0)[c]{$\CF^{6|12}$}}
\put(7.0,18.0){\makebox(0,0)[c]{$$}}
\put(55.0,18.0){\makebox(0,0)[c]{$$}}
\put(25.0,25.0){\vector(-1,-1){18}}
\put(37.0,25.0){\vector(1,-1){18}}
\end{picture}
\end{aligned}
\end{equation}
where $\CF^{6|12}:=\IC^{4|12}\times\IC P^1\times\IC P^1_*$ 
is called {\it correspondence space} and $\IC^{4|12}$ is 
complexified $\CN=3$ Minkowski superspace appearing as 
the $(4|12)$-dimensional moduli space of global holomorphic
sections of the bundle $\CL^{5|6}\to\IC P^1\times\IC P^1_*$
\cite{Witten:1978xx,Manin,Harnad:1988rs,Ward}. 
Furthermore, there is a one-to-one correspondence between 
gauge equivalence classes of solutions to the $\CN=3$ SYM 
equations in four dimensions and equivalence classes of 
holomorphic vector bundles\footnote{In the Dolbeault picture, 
these classes are described by gauge equivalence classes of 
solutions to the equations of motion of hCS theory on 
$\CL^{5|6}$.} $\CE$ over the quadric $\CL^{5|6}$ such that 
the vector bundles $\CE$ are holomorphically trivial on 
each submanifold 
$Y_\tx\cong\IC P^1\times\IC P^1_*\hookrightarrow
\CL^{5|6}$ with $\tx\in\IC^{4|12}$ 
\cite{Witten:1978xx,Manin,Harnad:1988rs}. Vector  
bundles $\CE$ having such properties are sometimes called
$\IC^{4|12}$-trivial \cite{Manin}.

In this paper, we introduce a {\it generalized superambitwistor 
space} $\CL^{5|6}_{m,n}$ via the short exact sequence
\begin{equation}\label{eq:seq2}
0\ \to\ \CL^{5|6}_{m,n}\ \to\ \CC^{6|6}_{m,n}\ 
  \overset{\kappa_{m,n}}{\to}\ 
  \CO(m,n)\ \to\ 0, 
\end{equation}
where the holomorphic vector bundles 
\begin{equation}\label{eq:defofcc}
 \CC^{6|6}_{m,n}\ :=\ 
     [\CO(m,n-1)\oplus\CO(m-1,n)]\otimes\IC^2\oplus
       [\Pi\CO(m,0)\oplus\Pi\CO(0,n)]\otimes\IC^3
\end{equation}
and
\begin{equation}
 \CO(m,n)\ :=\ {\rm pr}^*_1\CO(m)\otimes{\rm pr}^*_2\CO(n) 
\end{equation}
are fibred over $\IC P^1\times\IC P^1_*$. Here, the map 
$\kappa_{m,n}$ is defined as
\begin{equation}\label{eq:defkap}
 \kappa_{m,n}\,:\,(z^\a,w^\ad,\pi_\ad,\rho_\a,\eta_i,
   \te^i)\ 
     \mapsto\ z^\a\rho_\a-w^\ad\pi_\ad+2\te^i\eta_i,
\end{equation}
where $(z^\a,w^\ad,\pi_\ad,\rho_\a,\eta_i,\te^i)$ are 
homogeneous
coordinates on $\CC^{6|6}_{m,n}$. Clearly, for $m=n=1$ 
the sequence \eqref{eq:seq2} reduces to \eqref{eq:seq1}, 
i.e. $\CL^{5|6}\equiv\CL^{5|6}_{1,1}$ and
${\rm pr}_1^*\CP^{3|3}\oplus{\rm pr}_2^*\CP^{3|3}_*
\equiv\CC^{6|6}_{1,1}$.

One of the goals of our paper is to establish a 
supertwistor correspondence involving the space 
$\CL^{5|6}_{m,n}$. We first
discuss its geometry and the associated double fibration
\begin{equation}\label{eq:DF4}
\begin{aligned}
\begin{picture}(50,40)
\put(0.0,0.0){\makebox(0,0)[c]{$\CL^{5|6}_{m,n}$}}
\put(64.0,0.0){\makebox(0,0)[c]{$\IC^{M|N}$}}
\put(34.0,33.0){\makebox(0,0)[c]{$\CF^{M+2|N}$}}
\put(7.0,18.0){\makebox(0,0)[c]{$\pi_2$}}
\put(55.0,18.0){\makebox(0,0)[c]{$\pi_1$}}
\put(25.0,25.0){\vector(-1,-1){18}}
\put(37.0,25.0){\vector(1,-1){18}}
\end{picture}
\end{aligned}
\end{equation}
with $\CF^{M+2|N}:=\IC^{M|N}\times\IC P^1\times\IC P^1_*$ 
and
\begin{equation}\label{eq:genspti2}
 \IC^{M|N}\ :=\ \IC^{3mn+m+n-1|3(m+n+2)}\ \cong\ 
      \IC^{4|12}\times\IC^{3mn+m+n-5|3(m+n-2)}
\end{equation}
containing $\IC^{4|12}$ as a subspace. Afterwards, we will 
describe a one-to-one correspondence between the moduli 
space of solutions to SYM-type equations on the space 
\eqref{eq:genspti2} and the moduli space
of $\IC^{M|N}$-trivial holomorphic vector bundles over 
$\CL^{5|6}_{m,n}$. We shall call the obtained 
Yang-Mills-Higgs type equations
on $\IC^{M|N}$ the SYM hierarchy truncated up to level 
$(m,n)$. In addition, we show that the $\CN=3$ SYM 
 equations are contained in this system of differential 
equations. The full SYM hierarchy is introduced as an
asymptotic limit $n\to\infty$ after putting 
$m=n$.

Further, we show that the SYM hierarchy is 
associated with an infinite-dimensional 
algebra,
denoted by
$T^{4|12}[\lambda,\mu]$, with generators 
\begin{equation}\label{eq:generators}   
 \mu^a\lambda^{\dot b}P_{\a\bd},\qquad
\mu^a Q_{i\a}\qquad{\rm and}\qquad\lambda^{\dot b} Q^i_\bd
\end{equation}
for $a,\dot b\in\IN$. Here, $\lambda$ and $\mu$ are
local coordinates on $\IC P^1\times\IC P^1_*$ and
$P_{\a\bd}$, $Q_{i\a}$ and
$Q^i_{\ad}$ are the generators of the algebra
$T^{4|12}$ of supertranslations on
$\IC^{4|12}$. We have
\begin{equation}
 \big\{\mu^a Q_{i\a},\lambda^{\dot b}Q^j_\bd\big\}\ =\ 
-2\d^j_i\mu^a\lambda^{\dot b}P_{\a\bd}
\end{equation}
as the only nonvanishing (anti)commutation relations of
$T^{4|12}[\lambda,\mu]$.
 For a Minkowski metric on space-time 
and proper reality conditions on the extra coordinates,
we will
 obtain a corresponding real form of the above algebra. 
Notice that $T^{4|12}[\lambda,\mu]$
is a subalgebra of an algebra
 $T^{4|12}[\lambda,\lambda^{-1},\mu,\mu^{-1}]$ with the same 
generators \eqref{eq:generators} but for $a,\dot b\in\IZ$.
In addition, we 
demonstrate that $T^{4|12}[\lambda,\mu]$ generates hidden 
symmetries of the $\CN=3$ SYM equations. In this respect,
we again emphasize that the $\CN=3$ and $\CN=4$ SYM theories
are actually equivalent, so we will not make
any distinction between these theories and also refer to 
them interchangeably. The 
symmetries under consideration
are in fact point symmetries of the SYM 
hierarchy and the dependence of the space-time 
components of the SYM fields on the moduli 
parametrizing higher flows can be recovered by
solving a part of the infinite set of equations 
of the SYM hierarchy. In other
words, the generators of the algebra 
$T^{4|12}[\lambda,\mu]$ are
realized as (super)derivatives along 
``higher times" and they generate
tangent vectors to the infinite-dimensional 
space of solutions to
the $\CN=4$ SYM equations. The existence of 
such symmetries could
play an important role in quantum integrability 
of $\CN=4$ SYM theory.

\section{Generalized superambitwistor space}

\paragraph{Moduli space.} 
To clarify the geometry of $\CL^{5|6}_{m,n}$,
we use the exact sequence \eqref{eq:seq2}. 
This sequence induces
a long exact cohomology sequence
\begin{equation}\label{eq:cohomseq}
 \begin{aligned}
    0\ &\to\ H^0(Y,\CL^{5|6}_{m,n})\ \to\
  H^0(Y,\CC^{6|6}_{m,n})\ \overset{\kappa_{m,n}}{\to}\ 
          H^0(Y,\CO(m,n))\ \to\ \\ 
        &\kern.5cm\to\  H^1(Y,\CL^{5|6}_{m,n})\ \to\ 
          H^1(Y,\CC^{6|6}_{m,n})\ \to\ 
          H^1(Y,\CO(m,n))\ \to\ \cdots,
 \end{aligned}
\end{equation}
where $Y:=\IC P^1\times\IC P^1_*$ and the map 
$\kappa_{m,n}$ is defined by the formula 
\eqref{eq:defkap}.\footnote{By a slight abuse of 
notation, we use the same symbol $\kappa_{m,n}$ for both 
the map between vector bundles and the
map between cohomology groups.} 
Using K\"unneth's formula, we have
$H^1(Y,\CC^{6|6}_{m,n})=0=H^1(Y,\CO(m,n))$ for $m,n\geq1$. 
Furthermore, the map $\kappa_{m,n}$ is surjective. 
Therefore, the 
sequence \eqref{eq:cohomseq} yields a short exact sequence
\begin{equation}\label{eq:sseq1}
0\ \to\ H^0(Y,\CL^{5|6}_{m,n})\ \to\
          H^0(Y,\CC^{6|6}_{m,n})\
    \to\ H^0(Y,\CO(m,n))\ \to\ 0.
\end{equation}
In addition, we obtain
\begin{equation}\label{eq:sseq2}
 H^0(Y,\CC^{6|6}_{m,n})\ =\ \IC^{4mn+2m+2n|3(m+n+2)},
  \qquad
 H^0(Y,\CO(m,n))\ =\ \IC^{(m+1)(n+1)}
\end{equation}
and from \eqref{eq:sseq1} and \eqref{eq:sseq2} we find
\begin{equation}\label{eq:sseq3}
 H^0(Y,\CL^{5|6}_{m,n})\ =\ \IC^{M|N}\quad{\rm with}\quad
 M\ :=\ 3mn+m+n-1\quad{\rm and}\quad N\ :=\ 3(m+n+2).
\end{equation}
Thus, we conclude that the moduli space of global 
holomorphic sections of the
fibration
\begin{equation}\label{eq:sseq4}
 \CL^{5|6}_{m,n}\ \to\ Y
\end{equation}
is the $(M|N)$-dimensional superspace $\IC^{M|N}$ with 
$M,N$ given
in \eqref{eq:sseq3}.

\paragraph{Decomposition of $\IC^{M|N}$.} 
Let us consider the bosonic
part of the sequence \eqref{eq:seq2},
\begin{equation}
 0\ \to\ \CL^{5|0}_{m,n}\ \to\ 
 [\CO(m,n-1)\oplus\CO(m-1,n)]\otimes\IC^2\ \to\ 
 \CO(m,n)\ \to\ 0.
\end{equation}
Upon dualizing it and using the Euler sequence 
(see e.g. \cite{Griffith}), we conclude that 
$\CL^{5|0}_{m,n}$ can be identified with the
vector bundle Jet$^1\CO(m,n)$ of first order jets 
of $\CO(m,n)$. Moreover, by recalling the exact 
sequence (see e.g. \cite{Manin})
\begin{equation}\label{eq:jet}
 0\ \to\ T^*X\otimes E\to\ {\rm Jet}^1E\ \to\ E\ \to\ 0,
\end{equation}
 where $E$ is a vector bundle over a manifold $X$, 
we obtain
\begin{equation}\label{eq:seq3}
 0\ \to\ \CO(m-2,n)\oplus\CO(m,n-2)\ \to\ 
 \CL^{5|0}_{m,n}\ \to\ \CO(m,n)\ \to\ 0,
\end{equation}
where we have substituted $E=\CO(m,n)$ and 
$T^*X=T^*Y=T^*(\IC P^1\times\IC P^1_*)\cong
\CO(-2,0)\oplus\CO(0,-2)$ into \eqref{eq:jet}.

It is not difficult to show that the long exact 
cohomology sequence
arising from the short exact sequence \eqref{eq:seq3} 
reduces to
\begin{equation}\label{eq:seq4}
 0\ \to\ H^0(Y,\CO(m-2,n)\oplus\CO(m,n-2))\ 
\to\ H^0(Y,\CL^{5|0}_{m,n})\ \to\ 
 H^0(Y,\CO(m,n))\ \to\ 0.
\end{equation}
 As $H^0(Y,\CL^{5|0}_{m,n})=\IC^{M|0}$, 
\eqref{eq:seq4} implies the
decomposition
\begin{equation}
 \IC^{M|0}\ \cong\ \IC^{(m+1)(n+1)|0}
 \times\IC^{(m+1)(n-1)|0}\times
 \IC^{(m-1)(n+1)|0}
\end{equation}
of the bosonic part $\IC^{M|0}\subset\IC^{M|N}$ 
of the moduli space \eqref{eq:sseq3} of global 
holomorphic sections of the bundle
\eqref{eq:sseq4}. Using this natural 
decomposition, we choose the
following bosonic coordinates on $\IC^{M|N}$:
\begin{equation}\label{eq:coord1}
 x^{\a_1\cdots\a_n\ad_1\cdots\ad_m},\qquad
 t^{\a_1\cdots\a_{n-2}\ad_1\cdots\ad_m}
 \qquad{\rm and}\qquad
 s^{\a_1\cdots\a_n\ad_1\cdots\ad_{m-2}}.
\end{equation}
As fermionic coordinates, we may take
\begin{equation}\label{eq:coord2}
 \eta_i^{\ad_1\cdots\ad_m}\qquad{\rm and}\qquad 
 \te^{i\a_1\cdots\a_n}.
\end{equation}
Note that all of these coordinates are totally 
symmetric in their 
spinorial indices.

\paragraph{Geometry of the space $\CL^{5|6}_{m,n}$.}
As homogeneous coordinates on $\CC^{6|6}_{m,n}$ we take
$(z^\a,w^\ad,\pi_\ad,\rho_\a,\eta_i,$ $\te^i)\in\IC^{8|6}$ 
subject
to the identification
\begin{equation}\label{eq:identi}
 (z^\a,w^\ad,\pi_\ad,\rho_\a,\eta_i,\te^i)\ \sim\
 (t_1^m t_2^{n-1}z^\a,t_1^{m-1}t_2^n w^\ad,
  t_1\pi_\ad,t_2\rho_\a,t_1^m\eta_i,t_2^n\te^i)
\end{equation}
for any pair $(t_1,t_2)\in\IC^*\times\IC^*$ such that 
$(\pi_\ad)^t\neq(0,0)$ and
$(\rho_\a)^t\neq(0,0)$. Upon
imposing the quadric constraint 
\begin{equation}\label{eq:quadric}
 z^\a\rho_\a-w^\ad\pi_\ad+2\te^i\eta_i\ =\ 0
\end{equation}
on these coordinates, we may also use them as coordinates  
on $\CL^{5|6}_{m,n}$. Note that this constraint can be solved
on each of the four patches $\CW_p$, with 
$p,q,\ldots=1,\ldots,4$, covering 
$\CL^{5|6}_{m,n}$. Recall that $\CC^{6|6}_{m,n}$ is a 
holomorphic vector
bundle \eqref{eq:defofcc} over $Y=\IC P^1\times\IC P^1_*$ 
and as such it can be covered by four patches 
$\CU_p=\IC^{4|6}\times W_p$, where
$\{W_p\}$ is the standard acyclic covering of $Y$. 
We may choose $\CW_p=\CU_p\cap\CL^{5|6}_{m,n}$ since
$\CL^{5|6}_{m,n}\subset\CC^{6|6}_{m,n}$.

In terms of the homogeneous coordinates on $\CC^{6|6}_{m,n}$, 
global
holomorphic sections of the bundle \eqref{eq:sseq4} have 
the form
\begin{equation}\label{eq:holsec}
\begin{aligned}
 z^{\a_1}\ &=\ 
  [x^{\a_1\cdots\a_n\ad_1\cdots\ad_m}_R+\epsilon^{\a_1(\a_2}
   t^{\a_3\cdots\a_n)\ad_1\cdots\ad_m}]\rho_{\a_2}\cdots
     \rho_{\a_n}\pi_{\ad_1}\cdots\pi_{\ad_m},\\
     w^{\ad_1}\ &=\ [x^{\a_1\cdots\a_n\ad_1\cdots\ad_m}_L-
   s^{\a_1\cdots\a_n(\ad_3\cdots\ad_m}
        \epsilon^{\ad_2)\ad_1}]\rho_{\a_1}\cdots
         \rho_{\a_n}\pi_{\ad_2}\cdots\pi_{\ad_m},\\
    \eta_i\ &=\ \eta^{\ad_1\cdots\ad_m}_i
           \pi_{\ad_1}\cdots\pi_{\ad_m},\qquad
       \te^i\ =\ \te^{i\a_1\cdots\a_n}
             \rho_{\a_1}\cdots\rho_{\a_n}.
\end{aligned}
\end{equation}
Furthermore,
the moduli $x^{\a_1\cdots\a_n\ad_1\cdots\ad_m}_{R,L}$ 
appearing
in \eqref{eq:holsec} are not independent 
but satisfy
\begin{equation}
x^{\a_1\cdots\a_n\ad_1\cdots\ad_m}_R-
   x^{\a_1\cdots\a_n\ad_1\cdots\ad_m}_L
   +2\te^{i\a_1\cdots\a_n}\eta_i^{\ad_1\cdots\ad_m}\ =\ 0
\end{equation}
as a result of the quadric constraint \eqref{eq:quadric}. These
equations can be solved in terms of the coordinates on 
$\IC^{M|N}$ as
\begin{equation}
   x^{\a_1\cdots\a_n\ad_1\cdots\ad_m}_{R,L}\ =\
      x^{\a_1\cdots\a_n\ad_1\cdots\ad_m}
        \mp\g_{R,L}^{\a_1\cdots\a_n\ad_1\cdots\ad_m}
      \te^{i\a_1\cdots\a_n}\eta_i^{\ad_1\cdots\ad_m},
\end{equation}
where the constant factors 
$\g_{R,L}^{\a_1\cdots\a_n\ad_1\cdots\ad_m}$ obey
\begin{equation}
\g_R^{\a_1\cdots\a_n\ad_1\cdots\ad_m}+
       \g_L^{\a_1\cdots\a_n\ad_1\cdots\ad_m}\ =\ 2.
\end{equation}
Below, we shall choose them appropriately.

\paragraph{Double fibration.} 
Recall that the sections \eqref{eq:holsec}
describe embeddings 
$Y_\tx\hookrightarrow\CL^{5|6}_{m,n}$, where
$\tx$ collectively denotes all the coordinates 
$(x,t,s,\eta,\te)\in\IC^{M|N}$.
On the other hand, for each point 
$\ell=(z^\a,w^\ad,\pi_\ad,\rho_\a,\eta_i,\te^i)
\in\CL^{5|6}_{m,n}$, the {\it incidence relations} 
\eqref{eq:holsec} define
a subspace $\IC^{M-3|N-6}$ in $\IC^{M|N}$. 
The correspondence 
between subspaces in $\CL^{5|6}_{m,n}$ and 
$\IC^{M|N}$ can be 
described by the double fibration \eqref{eq:DF4} 
mentioned in the
previous section. There, $\CF^{M+2|N}=
\IC^{M|N}\times\IC P^1\times\IC P^1_*$
and the projections $\pi_1$ and $\pi_2$ are defined 
by the formulae
\begin{equation}
\begin{aligned}
    \pi_1\,:\,(x,t,s,\pi_\ad,\rho_\a,
               \eta_i^{\ad_1\cdots\ad_m},
         \te^{i\a_1\cdots\a_n})\ 
          &\mapsto\ (x,t,s,\eta_i^{\ad_1\cdots\ad_m},
              \te^{i\a_1\cdots\a_n}),\\
    \pi_2\,:\,(x,t,s,\pi_\ad,\rho_\a,
      \eta_i^{\ad_1\cdots\ad_m},\te^{i\a_1\cdots\a_n})\ 
      &\mapsto\ (z^\a,w^\ad,\pi_\ad,
            \rho_\a,\eta_i,\te^i),
\end{aligned}
\end{equation}
where $z^\a$, $w^\ad$, $\eta_i$ and $\te^i$ 
are given in \eqref{eq:holsec}.
So, the diagram \eqref{eq:DF4} describes 
the one-to-one correspondences
\begin{equation}
 \begin{tabular}{ccc}
    $\{$submanifolds $Y_\tx$ in $\CL^{5|6}_{m,n}\}
            $&$\quad\longleftrightarrow\quad$ & 
   $\{$points $\tx$ in $\IC^{M|N}\},$\\
     $\{$points $\ell$ in $\CL^{5|6}_{m,n}\}$ 
         & $\quad\longleftrightarrow\quad$ & $\{$subspaces 
             $\IC^{M-3|N-6}_\ell$ in $\IC^{M|N}$\},\\
 \end{tabular}
\end{equation}
where a fixed point $\tx\in\IC^{M|N}$ corresponds to a 
submanifold
$\pi_2(\pi_1^{-1}(\tx))\cong Y_\tx\hookrightarrow
\CL^{5|6}_{m,n}$ and, conversely, a
fixed point $\ell\in\CL^{5|6}_{m,n}$ corresponds 
to a codimension
$3|6$ subspace $\pi_1(\pi_2^{-1}(\ell))\cong
\IC^{M-3|N-6}_\ell
\hookrightarrow\IC^{M|N}$. Note that the correspondence 
space $\CF^{M+2|N}$ can be covered by four patches 
$\hCW_p:=\IC^{M|N}\times W_p$ with
local coordinates \eqref{eq:coord1}, \eqref{eq:coord2} 
on $\IC^{M|N}$ and
$(\lambda_{(p)},\mu_{(p)})$ on $W_p$, where $\{W_p\}$ 
forms the four-set acyclic covering
of $\IC P^1\times\IC P^1_*$.

\paragraph{Parametrization of $\IC^{M|N}$.} 
{}From now on, we specialize
to the case when $m=n$ and denote 
$\CL^{5|6}_n:=\CL^{5|6}_{n,n}$. The
discussion for $m\neq n$ can be given in an 
analogous way as for $m=n$.
Furthermore, we switch from the spinorial 
notation used above to a
polynomial one. This will allow us to write 
many formulae more
concisely and to also take the limit 
$n\to\infty$ in the field
equations. So, let us now consider the double 
fibration
\begin{equation}\label{eq:DF5}
\begin{aligned}
\begin{picture}(50,40)
\put(0.0,0.0){\makebox(0,0)[c]{$\CL^{5|6}_n$}}
\put(64.0,0.0){\makebox(0,0)[c]{$\IC^{M|N}$}}
\put(34.0,33.0){\makebox(0,0)[c]{$\CF^{M+2|N}$}}
\put(7.0,18.0){\makebox(0,0)[c]{$\pi_2$}}
\put(55.0,18.0){\makebox(0,0)[c]{$\pi_1$}}
\put(25.0,25.0){\vector(-1,-1){18}}
\put(37.0,25.0){\vector(1,-1){18}}
\end{picture}
\end{aligned}
\end{equation}
with $M=3n^2+2n-1$ and $N=6(n+1)$. In particular, 
we parametrize $\IC^{M|N}$ by the coordinates
\begin{equation}\label{eq:polcoord}
 (x^{A\dot B},t^{a\dot B},s^{A\dot b},
     \eta_i^{\dot A},\te^{i A}),
\end{equation}
where $A,B,\ldots=1,\ldots,n+1$,
$\dot A,\dot B,\ldots =\dot1,\ldots,\dot n+\dot1$ 
and $a,b,\ldots=1,\ldots,n-1$, 
$\dot a,\dot b,\ldots=\dot1,\ldots,\dot n-\dot1$. 
On the patch $\CU_1$ of $\CC_n^{6|6}:=\CC_{n,n}^{6|6}$ 
defined by the conditions $\pi_{\dot1}\neq 0$
and $\rho_1\neq0$, 
we may choose 
the following (local) coordinates:
\begin{equation}
\begin{aligned}
 z^\a_{(1)}\ :=\ \frac{z^\a}{\pi^n_{\dot1}\rho^{n-1}_1},
 \qquad
 z^3_{(1)}\ :=\ \frac{\pi_{\dot2}}{\pi_{\dot1}},\qquad
 w^\ad_{(1)}\ :=\ \frac{w^\ad}{\pi^{n-1}_{\dot1}\rho^n_1},
 \qquad
 w^{\dot3}_{(1)}\ :=\ \frac{\rho_2}{\rho_1},\\
 \eta^{(1)}_i\ :=\ \frac{\eta_i}{\pi^n_{\dot1}},\qquad
 \te^i_{(1)}\ :=\ \frac{\te^i}{\rho^n_1}.
\end{aligned}
\end{equation}
Note that by virtue of the projection $\pi_2$ 
appearing in \eqref{eq:DF5} we can set $z^3_{(1)}=\lambda_{(1)}$ 
and $w^{\dot3}_{(1)}=\mu_{(1)}$. Locally, the sections
\eqref{eq:holsec} over $W_1\subset Y$ then take the form\footnote{We
suppress (in most cases) the patch indices.}
\begin{equation}\label{eq:polsec}
\begin{aligned}
 z^1\ &=\ \sum_{\dot B=\dot1}^{\dot n+\dot1}
                   \left(x_R^{1\dot B}+\tfrac{1}{2}
   \sum_{a=1}^{n-1}(x^{a+1\dot B}_R+t^{a\dot B})\mu^a\right)
                  \lambda^{\dot B-\dot1},\\
       z^2\ &=\ \sum_{\dot B=\dot1}^{\dot n+\dot1}
       \left(\tfrac{1}{2}
      \sum_{a=1}^{n-1}(x^{a+1\dot B}_R-t^{a\dot B})\mu^{a-1}
                  +x_R^{n+1\dot B}\mu^{n-1}\right)
                  \lambda^{\dot B-\dot1},\\
       w^{\dot1}\ &=\ \sum_{A=1}^{n+1}\left(x_L^{A\dot1}+
             \tfrac{1}{2}
                  \sum_{\dot b=\dot1}^{\dot n-\dot1}
                    (x_L^{A\dot b+\dot1}+s^{A\dot b})
                  \lambda^{\dot b}\right)\mu^{A-1},\\
       w^{\dot2}\ &=\ \sum_{A=1}^{n+1}\left(\tfrac{1}{2}
    \sum_{\dot b=\dot1}^{\dot n-\dot 1}(x_L^{A\dot b+\dot1}-
                   s^{A\dot b})
      \lambda^{\dot b-\dot1}+x_L^{A \dot n+\dot 1}\lambda^{\dot n-\dot1}
                      \right)\mu^{A-1},\\
       \eta_i\ &=\ \sum_{\dot A=\dot1}^{\dot n+\dot 1}
                       \eta^{\dot A}_i
         \lambda^{\dot A-\dot1}\qquad{\rm and}\qquad
       \te^i\ =\ \sum_{A=1}^{n+1}\te^{iA}\mu^{A-1}.
\end{aligned}
\end{equation}
Similar expressions can also be written for the other patches 
$W_p$ with $p=2,3,4$. However, for illustrating purposes, 
we will mostly write 
formulae only for the patches $W_1$, $\CW_1$ and $\hCW_1$. 
As before, \eqref{eq:quadric} implies
\begin{equation}
x_R^{A\dot B}-x_L^{A\dot B}+2\te^{iA}\eta_i^{\dot B}\ =\ 0,
\end{equation}
which can be solved by putting
\begin{equation}
x_{R,L}^{A\dot B}\ =\ x^{A\dot B}\mp\g^{A\dot B}_{R,L}
       \te^{iA}\eta_i^{\dot B}
\end{equation}
with $\g_R^{A\dot B}+\g_L^{A\dot B}=2$. 
We shall specify these factors 
in a moment.

Among the coordinates \eqref{eq:polcoord} on $\IC^{M|N}$
 we will single
out those which correspond to coordinates on 
complexified $\CN=3$
Minkowski superspace $\IC^{4|12}$ entering in the 
decomposition
\begin{equation}\label{eq:decn}
 \IC^{M|N}\ =\ \IC^{3n^2+2n-1|6(n+1)}\ 
\cong\ \IC^{4|12}\times\IC^{3n^2+2n-5|6(n-1)}.
\end{equation}
The remaining coordinates are then interpreted as 
additional parameters (moduli) or
``higher times" from the viewpoint of $\CN=3$ SYM
 theory. The decomposition
\eqref{eq:decn} means the decomposition of the 
indices $(A)=(\a,a+2)$ and
$(\dot A)=(\ad,\dot a+\dot2)$ with $\a=1,2$, 
$a=1,\ldots,n-1$ and
$\ad=\dot1,\dot2$, $\dot a=\dot1,\ldots,\dot n-\dot1$. 
Using this, we denote
the coordinates on $\IC^{4|12}$ by
\begin{equation}
 (x^{\a\bd},\eta^\ad_i,\te^{i\a})
\end{equation}
and those on $\IC^{3n^2+2n-5|6(n-1)}$ by
\begin{equation}
 (x^{\a\,\dot b+\dot2},x^{a+2\,\bd},
x^{a+2\,\dot b+\dot2},t^{\a\dot b},t^{a+2\,\dot b},
 s^{a\bd},s^{a\,\dot b+\dot2},
\eta^{\dot a+\dot2}_i,\te^{ia+2}).
\end{equation}

\paragraph{Integrable distribution $\CT$.}
Below, we shall establish a correspondence 
between holomorphic vector 
bundles\footnote{In the Dolbeault picture, 
they can be described by hCS theory
on $\CL^{5|6}_n$.} over $\CL^{5|6}_n$ and 
solutions to the equations of
the SYM hierarchy truncated up to level $n$. 
Since the lowest level flows
of the hierarchy will -- by construction -- 
correspond to supertranslations
on $\IC^{4|12}$, the 
$\CN=3$ SYM equations are embedded
into the SYM hierarchy. Moreover, the set of vector 
fields spanning the
tangent spaces to the leaves of the fibration
\begin{equation}\label{eq:afib}
 \CF^{M+2|N}\ \to\ \CL^{5|6}_n
\end{equation}
must contain the vector fields
\begin{equation}\label{eq:VFSYM}
 \lambda\mu\der{x^{1\dot1}}-\mu\der{x^{1\dot2}}-
 \lambda\der{x^{2\dot1}}+\der{x^{2\dot2}},
 \quad \mu D_{i1}-D_{i2}\quad{\rm and}
 \quad\lambda D^i_{\dot1}-D^i_{\dot2}
\end{equation}
with
\begin{equation}\label{eq:fermder1}
 D_{i\a}\ :=\ \der{\te^{i\a}}+\eta_i^\bd\der{x^{\a\bd}}
  \qquad{\rm and}\qquad
 D^i_\ad\ :=\ \der{\eta^\ad_i}+\te^{i\b}\der{x^{\b\ad}}
\end{equation}
 entering into the linear system for the 
$\CN=3$ SYM equations. 
To implement this, we need to choose the factors 
$\g_{R,L}^{A\dot B}$
such that
\begin{equation}
\begin{aligned}
x_{R,L}^{\a\bd}\ &=\ x^{\a\bd}\mp\te^{i\a}\eta^{\bd}_i,\\
        x_R^{\a\,\dot b+\dot2}\ &=\ x^{\a\,\dot b+\dot2},\qquad
        x_L^{\a\,\dot b+\dot2}\ =\ x^{\a\,\dot b+\dot2} +2\te^{i\a}
           \eta^{\dot b+\dot2}_i,\\
  x_R^{a+2\,\bd}\ &=\ x^{a+2\,\bd}-2\te^{ia+2}\eta^{\bd}_i,\qquad
        x_L^{a+2\,\bd}\ =\ x^{a+2\,\bd},\\
  x_{R,L}^{a+2\,\dot b+\dot2}\ &=\ x^{a+2\,\dot b+\dot2}\mp\te^{ia+2}
        \eta^{\dot b+\dot2}_i.
\end{aligned}
\end{equation}

In the sequel, we denote the relative holomorphic tangent 
bundle of the
fibration \eqref{eq:afib} by 
$\CT$.\footnote{We shall use the same letter $\CT$ for both 
the bundle
and the distribution generated by its sections.} It is defined 
by the
short exact sequence
\begin{equation}
 0\ \to\ \CT\ \to\ T\CF^{M+2|N}\ \to\ 
\pi_2^*T\CL^{5|6}_n\ \to\ 0.
\end{equation}
In other words, by dualizing this sequence, we obtain the
sheaf $\Omega^1(\CF^{M+2|N})/\pi^*_2\Omega^1(\CL^{5|6}_n)$ of 
relative differential one-forms on $\CF^{M+2|N}$. Short 
calculations
reveal that for $n>1$
the bosonic part of the distribution $\CT$ is spanned by
the vector fields
\begin{subequations}\label{eq:TVF1}
\begin{eqnarray}
 D_{A\dot B} &=& \lambda\mu\der{x^{A\dot B}}-
          \mu\der{x^{A\dot B+\dot1}}-\lambda
         \der{x^{A+1\dot B}}+
            \der{x^{A+1\dot B+\dot1}}\notag\\ 
    &=&\mu^\a\lambda^\bd\der{x^{A+\a-1\,\dot B+\bd-\dot1}}\ =:\
         \mu^\a\lambda^\bd D_{\a\bd A\dot B},\label{eq:TVF1a}\\
 D_{n+1\dot A}&=&\lambda\left(\der{x^{1\dot A}}-\d^{\dot n}_{\dot A}
  \der{s^{1\,\dot n-\dot1}}\right)-\left(\der{x^{1\dot A+\dot 1}}+
       \d^{\dot1}_{\dot A}\der{s^{1\dot1}}\right)\ =:\ 
 \lambda^\ad D_{n+1\dot A\ad},\label{eq:TVF1b}\\
 D_{A\dot n+\dot1}&=&\mu\left(\der{x^{A\dot1}}-
                \d^n_A\der{t^{n-1\,\dot1}}\right)
  -\left(\der{x^{A+1\,\dot1}}+\d^1_A\der{t^{1\dot1}}\right)\ =:\
        \mu^\a D_{\a A\dot n+\dot1},\label{eq:TVF1c}\\
 T_{a\dot B}&=&\lambda\der{t^{a\dot B}}-\der{t^{a\,\dot B+\dot1}}\ =\ 
     \lambda^\bd\der{t^{a\,\dot B+\bd-\dot1}}\ =:\ 
\lambda^\bd D_{a\dot B\bd}\quad{\rm for}\quad
       a\ \leq\ n-1,\label{eq:TVF1d}\\
 T_{a\dot n+\dot1}&=&\mu\der{t^{a\,\dot1}}-\der{t^{a+1\,\dot1}}\ =\ 
        \mu^\a\der{t^{a+\a-1\,\dot1}}\ =:\ \mu^\a 
        D_{\a a\dot n+\dot1}\quad{\rm for}\quad
       a\ \leq\ n-2,\label{eq:TVF1e}\\
 S_{A\dot b}&=&\mu\der{s^{A\dot b}}-\der{s^{A+1\,\dot b}}\ =\ 
\mu^\a\der{s^{A+\a-1\,\dot b}}
     \ =:\ \mu^\a D_{\a A\dot b}\quad{\rm for}\quad
       \dot b\ \leq\ \dot n-\dot1,\label{eq:TVF1f}\\ 
 S_{n+1\dot a}&=&\lambda\der{s^{1\,\dot a}}-\der{s^{1\,\dot a+\dot1}}\ =\
           \lambda^\ad\der{s^{1\,\dot a+\ad-\dot1}}\ =:\
         \lambda^\ad D_{n+1\dot a\ad} \quad{\rm for}\quad
       \dot a\ \leq\ \dot n-\dot2,\label{eq:TVF1g}
\end{eqnarray}
\end{subequations}
where $A,B,\ldots\leq n$ and $\dot A,\dot B,\ldots\leq\dot n$ 
and
\begin{equation}
 (\lambda^\ad)\ :=\ \binom{\lambda}{-1}\qquad{\rm and}\qquad
 (\mu^\a)\ :=\ \binom{\mu}{-1}.
\end{equation}
For $n=1$, the bosonic part of $\CT$ is spanned by the vector field
$D_{1\dot1}=:D$ from \eqref{eq:TVF1a}.

In addition, the fermionic part of the distribution $\CT$ 
is spanned by the vector
fields
\begin{equation}\label{eq:TVF2}
 V_{iA}\ =\ \mu D_{iA}-D_{iA+1}\ =:\ \mu^\a D_{i\a A}
\qquad{\rm and}\qquad
 V^i_{\dot A}\ =\ \lambda D^i_{\dot A}-D^i_{\dot A+\dot 1}\ =:\ 
\lambda^\ad D^i_{\ad\dot A},\\
\end{equation}
where $D_{i\a}$ and $D^i_\ad$ were already given in 
\eqref{eq:fermder1}
and
\begin{equation}\label{eq:fermder2}
\begin{aligned}
 D_{ia+2}\ &=\ \der{\te^{ia+2}}+2\eta^\bd_i\der{x^{a+2\,\bd}}+
 \eta^{\dot b+\dot 2}_i\der{x^{a+2\,\dot b+\dot2}},\\
 D^i_{\dot a+\dot2}\ &=\ \der{\eta^{\dot a+\dot2}_i}+
        2\te^{i\b}\der{x^{\b\,\dot a+\dot2}}
     +\te^{ib+2}\der{x^{b+2\,\dot a+\dot2}},
\end{aligned}
\end{equation}
for $a,b\leq n-1$ and $\dot a,\dot b\leq\dot n-\dot1$. 
Besides the expressions
\eqref{eq:TVF1} and \eqref{eq:TVF2} on $\hCW_1$, one can easily 
write down
the basis vector fields of $\CT$ on the remaining patches 
$\hCW_2$,
$\hCW_2$ and $\hCW_4$ covering together with $\hCW_1$ the 
correspondence
space $\CF^{M+2|N}$. 

The only nonvanishing (anti)commutators among the above 
vector fields are
\begin{equation}
 \big\{V_{iA},V^j_{\dot B}\big\}\ =\ 2\d^j_i D_{A\dot B}.
\end{equation}
Hence, the distribution $\CT$ is integrable. To homogenize 
notation,
we define
\begin{equation}\label{eq:TVF3}
 D_{I'}\ :=\ \mu^\a\lambda^\bd D_{\a\bd I'},\qquad
 D_{\dot I}\ :=\ \mu^\a D_{\a\dot I}\qquad{\rm and}\qquad
 D_I\ :=\ \lambda^\ad D_{I\ad}
\end{equation}
with
\begin{equation}\label{eq:TVF4}
\begin{aligned}
 \{D_{I'}\}\ =\ \{D_{A\dot B}\}\qquad&{\rm and}\qquad 
\{D_{\a\bd I'}\}\ =\ 
       \{D_{\a\bd A\dot B}\},\\
 \{D_{\dot I}\}\ =\ \{S_{A\dot b},T_{a\dot n+\dot1},
 D_{A\dot n+\dot1}\}
   \qquad&{\rm and}\qquad \{D_{\a\dot I}\}\ =\
 \{D_{\a A\dot b},D_{\a a\dot n+\dot1},
         D_{\a A\dot n+\dot1}\},\\
 \{D_I\}\ =\ \{T_{a\dot B},S_{n+1\dot a},D_{n+1\dot A}\}
\qquad&{\rm and}\qquad
 \{D_{I\ad}\}\ =\ \{D_{a\dot B\ad},D_{n+1\dot a\ad},
D_{n+1\dot A\ad}\}. 
\end{aligned}
\end{equation}
Also, from \eqref{eq:TVF2} we see that
\begin{equation}
 D_{i\a A}\ =\ D_{iA+\a-1}\qquad{\rm and}\qquad
 D^i_{\ad\dot A}\ =\ D^i_{\dot A+\ad-\dot1},
\end{equation}
i.e. the derivatives $D_{i1A}$ and $D_{i2A}$, 
$D^i_{\dot1\dot A}$ and 
$D^i_{\dot 2\dot A}$ are not linearly independent. 
Similarly,  
\eqref{eq:TVF1}, \eqref{eq:TVF3} and \eqref{eq:TVF4} imply 
that for $n>1$
the derivatives $D_{\a\bd I'}$, $D_{\a\dot I}$ and $D_{I\ad}$ 
are not
independent. For $n=1$, there is only one bosonic and
six fermionic vector fields given in \eqref{eq:VFSYM}.

\paragraph{$\CN=3$ SYM equations.}
{}From the above formulae we see that 
$D_{1\dot1}$,
$V_{i1}$ and $V^i_{\dot1}$ coincide with the vector fields 
\eqref{eq:VFSYM} entering into
the linear system \cite{Witten:1978xx}
\begin{equation}\label{eq:LSSYM}
\begin{aligned}
 \mu^\a\lambda^\bd(\partial_{\a\bd}+\CA_{\a\bd})\psi_1\ &=:\ 
\mu^\a\lambda^\bd\nabla_{\a\bd}\psi_1\ =\ 0,\\
 \mu^\a(D_{i\a}+\CA_{i\a})\psi_1\ &=:\ 
         \mu^\a\nabla_{i\a}\psi_1\ =\ 0,\\
 \lambda^\ad(D^i_\ad+\CA^i_\ad)\psi_1\ &=:\ 
             \lambda^\ad\nabla^i_\ad\psi_1\ =\ 0,
\end{aligned}
\end{equation}
where $\partial_{\a\bd}:=\partial/\partial x^{\a\bd}$ and
$\psi_1$ is a matrix-valued function on $\hCW_1$ 
depending holomorphically
on $\lambda$ and $\mu$. The compatibility conditions of
 \eqref{eq:LSSYM}
are the constraint equations
\begin{equation}\label{eq:CE}
   \big\{\nabla_{i(\a},\nabla_{j\b)}\big\}\ =\ 0,\quad
    \big\{\nabla^i_{(\ad},\nabla^j_{\bd)}\big\}\ =\ 0
       \quad{\rm and}\quad
   \big\{\nabla_{i\a},\nabla^j_\bd\big\}
          -2\d^j_i\nabla_{\a\bd}\ =\ 0
\end{equation}
of $\CN=3$ SYM theory. Recall that these equations 
are equivalent to the 
field equations of $\CN=3$ SYM theory \cite{Witten:1978xx,HarnadVK}. 
Therefore, we will not make any distinction
between the constraint and the field equations.

\section{General form of the SYM hierarchy}

\paragraph{Relative connection $\CA_\CT$.}
Let $X$ be a smooth (super)manifold and $\CT$ an 
integrable distribution
on $X$. For any smooth function $f$ on $X$, let d$_\CT f$ be the 
restriction
of d$f$ to $\CT$, i.e. d$_\CT$ is the composition
\begin{equation}\label{eq:rd1}
C^\infty(X)\ \overset{\dt}{\to}\ \Omega^1(X)\
\to\ \Gamma(X,\CT^*),
\end{equation}
where $\Omega^1(X):=\Gamma(X,T^* X)$ and $\CT^*$ denotes the sheaf
of (smooth) differential one-forms dual to $\CT$. 
The operator
$\dt_\CT$ can be extended to act on relative differential
$k$-forms from the
space $\Omega^k_\CT(X):=\Gamma(X,\Lambda^k\CT^*)$,
\begin{equation}\label{eq:rd2}
\dt_\CT\, :\, \Omega^k_\CT(X)\ \to\
\Omega_\CT^{k+1}(X)\qquad{\rm and}\qquad k\ \geq\ 0.
\end{equation}

Let $\hat\CE$ be a smooth complex
vector bundle over $X$. A covariant differential (or connection)
on $\hat\CE$ along the distribution $\CT$ -- a $\CT$-connection
\cite{Rawnsley} -- is a $\IC$-linear mapping
\begin{equation}
D_\CT\, :\, \Gamma(X,\hat\CE)\ \rightarrow\ \Gamma(X,\CT^*\otimes\hat\CE)
\end{equation}
satisfying the Leibniz formula
\begin{equation}
D_\CT(f\sigma)\ =\ fD_\CT \sigma+\dt_\CT f\otimes \sigma,
\end{equation}
where $\sigma\in\Gamma(X,\hat\CE)$ is a local section of $\hat\CE$ and $f$
is a local smooth function. This $\CT$-connection extends to a map
\begin{equation}
D_\CT\, :\, \Omega^k_\CT(X,\hat\CE)\ \to\
\Omega_\CT^{k+1}(X,\hat\CE),
\end{equation}
where $\Omega^k_\CT(X,\hat\CE):=\Gamma(X,\Lambda^k\CT^*\otimes\hat\CE)$.
Locally, $D_\CT$ has the form
\begin{equation}
D_\CT\ =\ \dt_\CT+\CA_\CT,
\end{equation}
where the standard End$\,\hat\CE$-valued $\CT$-connection one-form
$\CA_\CT$ has components only along the distribution $\CT$. As
usual, $D^2_\CT$ naturally induces the curvature
$\CF_\CT=\dt_\CT\CA_\CT+\CA_\CT\wedge\CA_\CT\in
\Gamma(X,\Lambda^2\CT^*\otimes {\rm End}\,\hat\CE)$
of $\CA_\CT$. We say that $D_\CT$ (or
$\CA_\CT$) is flat, if $\CF_\CT=0$. For a flat $D_\CT$, the pair
$(\hat\CE,D_\CT)$ is called a $\CT$-flat vector bundle
\cite{Rawnsley}. 

\paragraph{Linear system on $\CF^{M+2|N}$.} 
Consider $X=\CF^{M+2|N}$ and
the integrable distribution $\CT$ which is locally spanned 
by the
vector fields \eqref{eq:TVF1} and \eqref{eq:TVF2}. We have
\begin{equation}\label{eq:rd3}
 \CT\ =\ {\rm span}\big\{ D_{I'},D_{\dot I},D_I,V_{iA},V^i_{\dot A}
          \big\}
\end{equation}
in the shorthand notation introduced in 
\eqref{eq:TVF2}, \eqref{eq:TVF3}
and \eqref{eq:TVF4}. Suppose we are given a topologically trivial
and $\IC^{M|N}$-trivial rank $r$ holomorphic vector bundle $\CE$ 
over
the generalized superambitwistor space $\CL^{5|6}_n$ from the 
double
fibration \eqref{eq:DF5}. In particular, this means that $\CE$ is
holomorphically trivial on any submanifold 
$Y_\tx\hookrightarrow\CL^{5|6}_n$
with $\tx\in\IC^{M|N}$. Let us consider the pulled-back bundle
$\tilde\CE:=\pi^*_2\CE$ over the supermanifold $\CF^{M+2|N}$ with a
covering $\{\hCW_p\}$ as defined in section 2. 
By definition, the pull-backs
$\tilde f:=\pi^*_2 f$ of the transition functions $f=\{f_{pq}\}$ of $\CE$
to $\tilde\CE=\pi^*_2\CE$ must be constant along the fibres of $\pi_2$,
i.e.\footnote{Recall that a (local) function on the 
correspondence space $\CF^{M+2|N}$ descends to generalized 
superambitwistor
space $\CL^{5|6}_n$ if and only if it lies in the kernel of 
$\dt_\CT$.}   
\begin{equation}\label{eq:condf}
 \dt_\CT\tilde f_{pq}\ =\ 0,
\end{equation}
where the relative differential $\dt_\CT$ (exterior differential
along the fibres of \eqref{eq:afib}) is defined by \eqref{eq:rd1},
\eqref{eq:rd2} and \eqref{eq:rd3}.

Note that $\IC^{M|N}$-triviality of the vector bundle 
$\CE\to\CL^{5|6}_n$ is 
equivalent to the triviality of the vector bundle 
$\tilde\CE\to\CF^{M+2|N}$
along the fibres $Y_\tx$ of the projection $\pi_1$ from 
\eqref{eq:DF5}.
Hence, there exist a trivialization $\{\psi_p\}$ of 
$\tilde\CE$ such that
\begin{equation}\label{eq:triv}
 \tilde f_{pq}\ =\ \psi^{-1}_p\psi_q\qquad{\rm on}\qquad
  \hCW_p\cap\hCW_q\ \neq\ \emptyset,
\end{equation}
where the $\psi_p$s are holomorphic in $\lambda_{(p)}$ and 
$\mu_{(p)}$. From
\eqref{eq:condf} and \eqref{eq:triv}, it follows that
\begin{equation}
 \psi_p\dt_\CT\psi_p^{-1}\ =\ \psi_q\dt_\CT\psi^{-1}_q
\end{equation}
on any $\hCW_p\cap\hCW_q$ and therefore
\begin{equation}
 \CA_\CT|_{\hCW_p}\ :=\ \psi_p\dt_\CT\psi_p^{-1}
\end{equation}
is a globally defined flat $\CT$-connection on a trivial 
vector bundle $\hat\CE\to\CF^{M+2|N}$ which is 
topologically equivalent to
$\tilde\CE$. In fact, we have an equivalence
\begin{equation}
 (\tilde\CE, \tilde f=\{\tilde f_{pq}\},\dt_\CT)\ \sim\
 (\hat\CE, \hat f=\{\mathbbm{1}_r\},D_\CT=\dt_\CT+\CA_\CT),
\end{equation}
which is the equivalence of the \v Cech and Dolbeault 
descriptions
of $\CT$-flat vector bundles.

By an extension of 
Liouville's theorem
to $Y\hookrightarrow\CF^{M+2|N}$, it follows that
on patch $\hCW_1$ the components of $\CA_\CT$ 
are given by
the formulae
\begin{equation}\label{eq:LSH1}
\begin{aligned}
 D_{I'}\lrcorner\CA_\CT\ &:=\ \psi_1 D_{I'}\psi_1^{-1}\ =\ 
    \mu^\a\lambda^\bd\CA_{\a\bd I'},\\
 D_{\dot I}\lrcorner \CA_\CT\ &:=\ \psi_1 D_{\dot I}\psi_1^{-1}
   \ =\ \mu^\a\CA_{\a\dot I},\\
 D_I\lrcorner\CA_\CT\ &:=\ \psi_1 D_I\psi_1^{-1}\ =\ 
    \lambda^\ad \CA_{I\ad},\\
 V_{iA}\lrcorner\CA_\CT\ &:=\ \psi_1 V_{iA}\psi_1^{-1}\ =\ 
 \mu^\a \CA_{i\a A},\\
 V^i_{\dot A}\lrcorner\CA_\CT\ &:=\ 
         \psi_1 V^i_{\dot A}\psi_1^{-1}\ =\
 \lambda^\ad \CA^i_{\ad\dot A},
\end{aligned}
\end{equation}
where the fields $\{\CA\}=\{\CA_{\a\bd I'},
\CA_{\a \dot I},\ldots\}$ do not depend
on $\lambda$ and $\mu$. Similar formulae can be written down for the 
other patches. Equations \eqref{eq:LSH1} can be rewritten as the
following system of linear differential equations:
\begin{equation}\label{eq:LSH2}
\begin{aligned}
 \mu^\a\lambda^\bd(D_{\a\bd I'}+\CA_{\a\bd I'})\psi_1\ &=:\ 
   \mu^\a\lambda^\bd\nabla_{\a\bd I'}\psi_1\ =\ 0,\\
 \mu^\a(D_{\a\dot I}+\CA_{\a\dot I})\psi_1\ &=:\ 
 \mu^\a\nabla_{\a\dot I}\psi_1\ =\ 0,\\
 \lambda^\ad(D_{I\ad}+\CA_{I\ad})\psi_1\ &=:\ 
 \lambda^\ad\nabla_{I\ad}\psi_1\ =\ 0,\\
 \mu^\a(D_{i\a A}+\CA_{i\a A})\psi_1\ &=:\
 \mu^\a\nabla_{i\a A}\psi_1\ =\ 0,\\
 \lambda^\ad(D^i_{\ad \dot A}+\CA^i_{\ad\dot A})\psi_1\ &=:\
 \lambda^\ad\nabla^i_{\ad\dot A}\psi_1\ =\ 0.
\end{aligned}
\end{equation}
It is not too difficult to see that these equations are invariant
under the gauge transformations
\begin{equation}\label{eq:GT}
 \psi_1\ \mapsto\ g^{-1}\psi_1\qquad{\rm and}\qquad
 \CA\ \mapsto\ g^{-1}\CA g+g^{-1}D g,
\end{equation}
where $g$ is any smooth $GL(r,\IC)$-valued function on $\IC^{M|N}$.
In addition, \eqref{eq:LSH2} is also invariant under
the equivalence map
\begin{equation}\label{eq:ET}
 \psi_1\ \mapsto\ \psi_1 h_1\qquad{\rm and}\qquad \CA\ \mapsto\ 
\CA,
\end{equation}
where $h_1$ is any holomorphic $GL(r,\IC)$-valued function on
$\hCW_1\subset\CF^{M+2|N}$. Similar invariance transformations hold 
for the remaining patches $\hCW_p$ for $p=2,3,4$. Note that the 
transition
functions \eqref{eq:triv} are invariant under the
transformations \eqref{eq:GT} and furthermore are mapped to 
equivalent 
ones for \eqref{eq:ET}.

\paragraph{SYM(n) equations.} 
The compatibility conditions of the equations \eqref{eq:LSH2} 
are given by
\begin{equation}\label{eq:compH}
\begin{aligned}
{\big[\nabla_{(\a(\ad I'},\nabla_{\b)\bd)J'}\big]}\ =\ 0,\quad
          \big[\nabla_{(\a\ad I'},\nabla_{\b)\dot J}\big]\ =\ 0,\quad
         \big[\nabla_{\a(\ad I'},\nabla_{J\bd)}\big]\ =\ 0,\\
         \big[\nabla_{(\a\ad I'},\nabla_{i\b)B}\big]\ =\ 0,\quad
         \big[\nabla_{\a(\ad I'},\nabla^i_{\bd)\dot B}\big]\ =\ 0,\\
         \big[\nabla_{(\a\dot I},\nabla_{\b)\dot J}\big]\ =\ 0,\quad
         \big[\nabla_{\a\dot I},\nabla_{J\bd}\big]\ =\ 0,\quad
          \big[\nabla_{I(\ad},\nabla_{J\bd)}\big]\ =\ 0,\\
       \big[\nabla_{(\a\dot I},\nabla_{i\b)B}\big]\ =\ 0,\quad
          \big[\nabla_{\a\dot I},\nabla^i_{\bd\dot B}\big]\ =\ 0,\quad
              \big[\nabla_{I\ad},\nabla_{i\b B}\big]\ =\ 0,\quad
        \big[\nabla_{I(\ad},\nabla^i_{\bd)\dot B}\big]\ =\ 0,\\
          \big\{\nabla_{i(\a A},\nabla_{j\b)B}\big\}\ =\ 0,\quad
           \big\{\nabla^i_{(\ad\dot A},\nabla^j_{\bd)\dot B}\big\}\ =\ 0,\quad
          \big\{\nabla_{i\a A},\nabla^j_{\bd\dot B}\big\}
               -2\d^j_i\nabla_{\a\bd A\dot B}\ =\ 0,
\end{aligned}
\end{equation}
where parentheses mean normalized symmetrization of the 
spinorial indices.

We shall call the finite system \eqref{eq:compH} of 
nonlinear differential equations 
the {\it SYM hierarchy truncated up to level $n$} and refer 
to the field equations
\eqref{eq:compH} as the {\it SYM(n) 
equations}. The full SYM hierarchy is 
obtained by taking
the asymptotic limit $n\to\infty$.\footnote{One can also allow the 
indices to run over
all negative values with corresponding coordinates $x^{A\dot B}$ 
(hence, $A,\dot B\in\IZ$),
etc. parametrizing ``negative flows".} We emphasize that because of 
Bianchi identities,
\eqref{eq:compH} is not the ``minimal" set of equations since some 
equations of
\eqref{eq:compH} are implied by others. As our subsequent discussion 
will not be
affected by this issue, we can leave it aside in the remainder of 
this work. 

It is not difficult to see that the constraint 
equations \eqref{eq:CE}
of $\CN=3$ SYM theory are embedded into the SYM($n$) equations. 
Indeed, we find them
\begin{equation}
 \big\{\nabla_{i(\a 1},\nabla_{j\b)1}\big\}\ =\ 0,\quad
\big\{\nabla^i_{(\ad\dot 1},\nabla^j_{\bd)\dot 1}\big\}\ =\ 0,
\quad
  \big\{\nabla_{i\a 1},\nabla^j_{\bd\dot 1}\big\}-
     2\d^j_i\nabla_{\a\bd 1\dot 1}\ =\ 0,
\end{equation}
as part of \eqref{eq:compH} since
$\nabla_{i\a1}\equiv\nabla_{i\a}$, $\nabla^i_{\ad\dot1}\equiv\nabla^i_\ad$
and $\nabla_{\a\bd1\dot1}\equiv\nabla_{\a\bd}$.

Now one could continue to work with \eqref{eq:compH} by imposing a sort
of transversal gauge \cite{HarnadVK} (or follow a more general approach
\cite{Arefeva:1986zx}) to extract the field content, the superfield
expansions, etc. of the SYM($n$) equations. However, already in 
describing truncated hierarchies for $\CN$-extended self-dual SYM theories,
this procedure was rather involving \cite{Wolf:2004hp}. 
That is why we leave
this discussion to future work. In the next section, we will write the 
SYM($n$) equations in light-cone gauge in which they take a simpler
form.

\paragraph{Reality conditions.}
Up to now, we have only worked in a complex setting. However, it is
well known that for $\CN=4$ SYM theory described in terms of
supertwistors, one can obtain real SYM fields on Minkowski space
$\IR^{1,3}$ with metric $g={\rm diag}(-1,+1,+1,+1)$ by choosing  
appropriate real structures on $\IC^{4|12}$, $\CP^{3|3}\times\CP^{3|3}_*$
and on the quadric $\CL^{5|6}$ (see e.g. \cite{Manin}). In fact,
by introducing an antiholomorphic involution on $\CL^{5|6}$, one
obtains an induced involution on $\IC^{4|12}$ such that its fixed point
set is identified with Minkowski superspace.

In the present case of generalized superambitwistor space, we
may proceed similarly. In particular, we consider the 
antiholomorphic
involution $\tau_M:\CL^{5|6}_n\to\CL^{5|6}_n$ which is defined
by its action on the homogeneous coordinates of $\CC^{6|6}_n$
as the map
\begin{equation}\label{eq:AHI}
 \tau_M\,:\, (z^\a,w^\ad,\pi_\ad,\rho_\a,\eta_i,\te^i)\ \mapsto\
            (-\overline{w^\ad},-\overline{z^\a},\overline{\rho_\a},
           \overline{\pi_\ad},
      \overline{\te^i},\overline{\eta_i}),
\end{equation}
 where bar denotes complex conjugation. Inserting
\eqref{eq:polsec} into \eqref{eq:AHI}, we find that $\tau_M$
acts on $\IC^{M|N}$ as\footnote{By a slight abuse of notation,
we use the symbol $\tau_M$ for maps defined on different spaces.}
\begin{equation}\label{eq:rcx1}
\tau_M(x^{A\dot B})\ =\ -\overline{x^{B\dot A}},\qquad
       \tau_M(t^{a\dot B})\ =\ -\overline{s^{B\dot a}}
          \qquad{\rm and}\qquad
       \tau_M(\eta^{\dot A}_i)\ =\ \overline{\te^{iA}}.
\end{equation}
Clearly, the fixed point set is a real slice 
$\IR^{M|N}\subset\IC^{M|N}$,
with 
\begin{equation}\label{eq:adec}    
\IR^{M|N}\ \cong\ \IR^{4|12}\times\IR^{M-4|N-12}\quad{\rm for}\quad 
 M\ =\ 3n^2+2n-1\quad{\rm and}\quad N\ =\ 6(n+1).
\end{equation}
If we choose the parametrization of $x^{\a\bd}$ according to
\begin{equation}\label{eq:rcx2}
 (x^{\a\bd})\ :=\ -{\rm i}
         \begin{pmatrix}
            x^0+x^3 & x^1-{\rm i} x^2\\
             x^1+{\rm i} x^2 & x^0-x^3
         \end{pmatrix}
\end{equation}
with $(x^0,x^1,x^2,x^3)\in \IR^4$,
the factor $\IR^{4|12}$ appearing in the decomposition 
\eqref{eq:adec}
can be identified with $\CN=3$ Minkowski superspace. 

The extension of
$\tau_M$ to matrix-valued holomorphic functions $h$ is given by
\begin{equation}
 \tau_M(h(\cdots))\ :=\ \overline{[h(\tau_M(\cdots))]}\,^t,
\end{equation}
where ``$t$" means matrix transposition.
With this rule we can extend
the involution $\tau_M$ to the holomorphic vector bundles $\CE$
and $\tilde\CE=\pi_2^*\CE$. In particular, for the transition functions
$\tilde f=\{\tilde f_{pq}\}$ of $\tilde\CE$ the reality condition
$\tau_M(\tilde f)=\tilde f$ yields
\begin{equation}\label{eq:RCtrans}
 \tilde f_{12}^\dagger\ =\ \tilde f_{31},\qquad
 \tilde f_{14}^\dagger\ =\ \tilde f_{41}\qquad{\rm and}\qquad
 \tilde f_{23}^\dagger\ =\ \tilde f_{23}.
\end{equation}
Here, we have used the shorthand notation
$\tilde f_{pq}^\dagger:=[\tilde f_{pq}
   (\tau_M(\cdots))]^\dagger$.
Equations \eqref{eq:RCtrans} can be satisfied by imposing the conditions
\begin{equation}\label{eq:RCpsi}
\psi_1^\dagger\ =\ \psi_1^{-1},\qquad
       \psi_2^\dagger\ =\ \psi_3^{-1}\qquad{\rm and}\qquad
       \psi_4^\dagger\ =\ \psi_4^{-1}
\end{equation}
which yield the relations
\begin{equation}
(\CA_{\a\bd A\dot B})^\dagger\ =\ \CA_{\b\ad B\dot A},\quad
    (\CA_{\a\dot I})^\dagger\ =\ \CA_{I\ad}\quad{\rm and}\quad
    (\CA^i_{\ad\dot A})^\dagger\ =\ \CA_{i\a A}.
\end{equation}
Therefore, the gauge group is reduced from $GL(r,\IC)$ to the unitary
group $U(r)$. If one in addition assumes that $\det(\tilde f_{pq})=1$,
then one obtains $SU(r)$.

To sum up, we have shown that there is a one-to-one correspondence
between $\IR^{M|N}$-trivial $\tau_M$-invariant holomorphic vector
bundles $\CE$ over the generalized superambitwistor space 
$\CL^{5|6}_n$ and gauge equivalence classes of solutions to the
equations of the truncated SYM($n$) hierarchy which include the
field equations of $\CN=3$ SYM theory as a subset.

\section{SYM hierarchy in light-cone gauge}

\paragraph{Extended linear system.}
For the sake of concreteness, we again work on the patch $\hCW_1$
of $\CF^{M+2|N}$. We start by reconsidering the vector fields
\eqref{eq:TVF1} and \eqref{eq:TVF2} and the resulting linear
system \eqref{eq:LSH2}. Note that the vector fields
\eqref{eq:TVF1a} are a linear combination of the vector fields
\begin{subequations}\label{eq:XY}
\begin{equation}\label{eq:X}
 X_{A\dot B}\ =\ \lambda\left(\der{x^{A\dot B}}-\d^{\dot n}_{\dot B}
 \der{s^{A\dot n-\dot1}}\right)-\left(\der{x^{A\dot B+\dot 1}}+
 \d^{\dot1}_{\dot B}\der{s^{A\dot1}}\right)\quad{\rm with}\quad
  A\ \leq\ n+1,\ \dot B\ \leq\ \dot n
\end{equation}
or
\begin{equation}\label{eq:Y}
Y_{A\dot B}\ =\ \mu\left(\der{x^{A\dot B}}-\d^n_A
 \der{t^{n-1\dot B}}\right)-\left(\der{x^{A+1\dot B}}+
 \d^1_A\der{t^{1\dot A}}\right)\quad{\rm with}\quad
  A\ \leq\ n,\ \dot B\ \leq\ \dot n+\dot1
\end{equation}
\end{subequations}
together with the remaining vector fields of \eqref{eq:TVF1}.
In particular, we have
\begin{subequations}\label{eq:somevf}
\begin{eqnarray}
 D_{A\dot B}&=&\mu X_{A\dot B}-
X_{A+1\dot B}+\d^{\dot1}_{\dot B} S_{A\dot 1}
         +\lambda\d^{\dot n}_{\dot B} S_{A\dot n-\dot1}\\
 &=& \lambda Y_{A\dot B}-Y_{A\dot B+\dot 1}+\d^1_A T_{1\dot B}+\mu
    \d^n_A T_{n-1\dot B}\qquad{\rm with}\qquad A\ \leq\ n,\ 
\dot B\ \leq\ \dot n.
\end{eqnarray}
\end{subequations}
In other words, both sets \eqref{eq:X} and \eqref{eq:Y} of 
vector fields
belong to the distribution $\CT$ tangent to the fibres of the 
projection
$\pi_2:\CF^{M+2|N}\to\CL^{5|6}_n$. Together with vector fields
\eqref{eq:TVF1d}--\eqref{eq:TVF1g} they form an ``overcomplete basis"
for the distribution $\CT$. However, they all do annihilate the 
transition
functions of the holomorphic vector bundle $\tilde\CE\to\CF^{M+2|N}$ 
and
their linear span forms $\CT$. Therefore, one can use them for 
introducing
an extended linear system\footnote{Of course, one now has to 
deal with
additional constraints among the obtained superfields caused by 
the linear
dependence of the used vector fields. However, this fact does not 
affect
our subsquent discussion.} which eventually yields a rather 
homogeneous
form of the equations of the SYM hierarchy when written in 
light-cone gauge.

Instead of \eqref{eq:TVF1a}--\eqref{eq:TVF1c} for $n>1$, we now 
take
the vector fields \eqref{eq:XY} and combine them together with 
the
remaining bosonic vector fields \eqref{eq:TVF1d}--\eqref{eq:TVF1g} 
into
the following expressions:
\begin{subequations}\label{eq:EVF}
\begin{equation}\label{eq:EVFa}
 D_{\dot\Upsilon}\ =\ \mu^\a D_{\a\dot \Upsilon}
\qquad{\rm and}\qquad
 D_\Upsilon\ =\ \lambda^\ad D_{\Upsilon\ad}, 
\end{equation}
where 
\begin{eqnarray}
 \{D_{\dot\Upsilon}\}\ =\ \{Y_{A\dot B},S_{A\dot b},
T_{a\dot n+\dot1}\}
   \quad&{\rm and}&\quad \{D_{\a\dot\Upsilon}\}\ =\ 
\{D_{\a A\dot B},D_{\a A\dot b},
         D_{\a a\dot n+\dot1}\},\label{eq:EVFb}\\
 \{D_\Upsilon\}\ =\ \{X_{A\dot B}, T_{a\dot B},S_{n+1\dot a}\}
\quad&{\rm and}&\quad
 \{D_{\Upsilon\ad}\}\ =\ \{D_{A\dot B\ad},D_{a\dot B\ad},
D_{n+1\dot a\ad}\}.\label{eq:EVFc} 
\end{eqnarray}
\end{subequations}
So, the abstract indices $\Upsilon$ and $\dot\Upsilon$ run over 
the appropriate
index set which is easily extracted from \eqref{eq:EVF}. 
Note that the
vector fields \eqref{eq:EVFb} and \eqref{eq:EVFc} are related 
by the involution
$\tau_M$.

Now, instead of the linear system \eqref{eq:LSH2}, we  
find the
following extended linear system:
\begin{equation}\label{eq:extLSH}
\begin{aligned}
 \mu^\a(D_{\a\dot\Upsilon}+\CA_{\a\dot\Upsilon})\psi_1\ &=\ 0,\\
 \lambda^\ad(D_{\Upsilon\ad}+\CA_{\Upsilon\ad})\psi_1\ &=\ 0,\\
 \mu^\a(D_{i\a A}+\CA_{i\a A})\psi_1\ &=\ 0,\\
 \lambda^\ad(D^i_{\ad\dot A}+\CA^i_{\ad\dot A})\psi_1\ &=\ 0.
\end{aligned}
\end{equation}

\paragraph{Light-cone gauge.}
Let us now fix a gauge by imposing the condition 
$\psi_1(\lambda=0=\mu)=1$. 
Such a gauge can always be obtained from the general 
$(\lambda,\mu)$-expansion,
\begin{equation}\label{eq:exp}
 \psi_1\ =\ \sum_{k,l=0}^\infty\lambda^k\mu^l\psi_1^{(k,l)},
\end{equation}
by performing the gauge transformation\footnote{Note that such
a transformation must be performed on all four patches 
simultaneously with
the same matrix $\psi^{(0,0)}_1$, i.e. 
$\psi_p\ \mapsto\ (\psi^{(0,0)}_1)^{-1}\psi_p$.} 
$\psi_1\mapsto\tilde\psi_1
=(\psi_1^{(0,0)})^{-1}\psi_1$ so that $\tilde\psi_1(\lambda=0=\mu)=1$. 
Hence,
from the very beginning we can assume 
that $\psi_1^{(0,0)}=1$ in \eqref{eq:exp}.
Then we have
\begin{equation}\label{eq:exp1}
\psi_1\ =\ 1+\mu\Phi+\lambda\Psi+\lambda\mu\Sigma+\cdots.
\end{equation}
Imposing the reality condition \eqref{eq:RCpsi} on $\psi_1$, we 
obtain
\begin{equation}\label{eq:RCTheta}
\Phi\ =\ -\Psi^\dagger\qquad{\rm and}\qquad
          \Sigma^\dagger\ =\ -\Sigma+\Psi\Phi+\Phi\Psi.
\end{equation}

Next, we substitute the expansion \eqref{eq:exp1} into the 
linear
system \eqref{eq:extLSH} and find that all the fields 
$\CA_{\a\dot\Upsilon}$, $\CA_{\Upsilon\ad}$, $\CA_{i\a A}$ and 
$\CA^i_{\ad\dot A}$ are expressed in terms of the prepotentials
$\Phi$, $\Psi$ and $\Sigma$ as
\begin{equation}\label{eq:pot}
\begin{aligned}
\CA_{1\dot\Upsilon}\ =\ D_{2\dot\Upsilon}\Phi,\qquad
\CA_{2\dot\Upsilon}\ =\ 0,\qquad
\CA_{\Upsilon\dot1}\ =\ D_{\Upsilon\dot2}\Psi,\qquad
\CA_{\Upsilon\dot2}\ =\ 0,\\
\CA_{i1A}\ =\ D_{i2A}\Phi,\qquad
\CA_{i2A}\ =\ 0,\qquad
\CA^i_{\dot1\dot A}\ =\ D^i_{\dot2\dot A}\Psi,\qquad
\CA^i_{\dot2\dot A}\ =\ 0. 
\end{aligned}
\end{equation}
In addition, the fields $\{\Phi,\Psi,\Sigma\}$ are constrained 
by the
differential equations
\begin{subequations}\label{eq:LCeq1}
\begin{eqnarray}
D_{\Upsilon\dot 2}\Phi\ =\ 0,\qquad
D^i_{\dot2\dot A}\Phi&=& 0,\label{eq:LCeq1b}\\
D_{2\dot\Upsilon}\Psi\ =\ 0,\qquad
D_{i2A}\Psi&=& 0,\label{eq:LCeq1a}\\
D_{\Upsilon\dot2}\Sigma-D_{\Upsilon\dot1}\Phi-
D_{\Upsilon\dot2}(\Psi\Phi)\ =\ 0,\quad
D_{2\dot\Upsilon}\Sigma-D_{1\dot\Upsilon}\Psi-
D_{2\dot\Upsilon}(\Phi\Psi)&=& 0,\label{eq:LCeq1c}\\
D^i_{\dot2\dot A}\Sigma-D^i_{\dot1\dot A}\Phi-
D^i_{\dot2\dot A}(\Psi\Phi)\ =\  0,\qquad
D_{i2A}\Sigma-D_{i1A}\Psi-D_{i2A}(\Phi\Psi)&=& 0.
\label{eq:LCeq1d}
\end{eqnarray}
\end{subequations}
Inserting \eqref{eq:pot} into the compatibility 
conditions of the linear
system \eqref{eq:extLSH} and using
\eqref{eq:LCeq1}, we obtain the equations
\begin{subequations}\label{eq:LCeq2}
\begin{eqnarray}
  D_{1\dot\Upsilon}D_{2\dot\Lambda}\Phi-
D_{1\dot\Lambda}D_{2\dot\Upsilon}\Phi
 +[D_{2\dot\Lambda}\Phi,D_{2\dot\Upsilon}\Phi]&=&0,\\
 D_{1\dot\Upsilon}D_{i2A}\Phi-D_{i1A}D_{2\dot\Upsilon}\Phi+
[D_{2\dot\Upsilon}\Phi,
  D_{i2A}\Phi]&=& 0,\label{eq:LCeq2b}\\
    D_{i1A}D_{j2B}\Phi+D_{j1B}D_{i2A}\Phi+
\{D_{i2A}\Phi,D_{j2B}\Phi\}&=&0,\\
  D_{\Upsilon\dot1}D_{\Lambda\dot2}\Psi-D_{\Lambda\dot1}
D_{\Upsilon\dot2}\Psi+
    [D_{\Upsilon\dot2}\Psi,D_{\Lambda\dot2}\Psi]&=&0,\\
  D_{\Upsilon\dot1}D^i_{\dot2\dot A}\Psi-D^i_{\dot1\dot A}
D_{\Upsilon\dot2}\Psi
  +[D_{\Upsilon\dot2}\Psi,D^i_{\dot2\dot A}\Psi]&=&0,\\
 D^i_{\dot1\dot A}D^j_{\dot2\dot B}\Psi+D^j_{\dot1\dot B}
D^i_{\dot2\dot A}\Psi
      +\{D^i_{\dot2\dot A}\Psi,D^j_{\dot2\dot B}\Psi\}&=&0,\\
 D_{1\dot\Upsilon}D_{\Lambda\dot2}\Psi-D_{\Lambda\dot1}
D_{2\dot\Upsilon}\Phi+
 [D_{2\dot\Upsilon}\Phi,D_{\Lambda\dot2}\Psi]&=&0,\\
 D_{1\dot\Upsilon}D^i_{\dot2\dot A}\Psi-D^i_{\dot1\dot A}
D_{2\dot\Upsilon}\Phi
 +[D_{2\dot\Upsilon}\Phi,D^i_{\dot2\dot A}\Psi]&=&0,\\
 D_{i1A}D_{\Upsilon\dot2}\Psi-D_{\Upsilon\dot1}D_{i2A}\Phi+
     [D_{i2A}\Phi,D_{\Upsilon\dot2}\Psi]&=&0,
\end{eqnarray}
\end{subequations}
which together with equations \eqref{eq:LCeq1} form the SYM($n$) 
equations (the
SYM hierarchy truncated up to level $n$) in light-cone gauge. As 
before, we obtain 
the full SYM hierarchy by taking the limit 
$n\to\infty$.\footnote{On may also
consider the case when all the indices run over all integers, 
i.e. over $\IZ$.}

\paragraph{$\CN=3$ SYM theory in light-cone gauge.}
Before continuing, we briefly explain why we called this gauge 
``light-cone gauge".
Recall that in terms of \eqref{eq:XY} the vector fields 
\eqref{eq:TVF1a} are given
by the formulae \eqref{eq:somevf}. Therefore, the gauge potential
$\CA_{\a\bd A\dot B}$ which is associated with $D_{\a\bd A\dot B}$ 
from
\eqref{eq:TVF1a} is represented as a certain
linear combination of the above introduced fields 
$\CA_{\a\dot\Upsilon}$ and
$\CA_{\Upsilon\ad}$ and, of course, similarly for
$\CA_{\a \dot I}$ and $\CA_{I\ad}$. A short calculation then shows that 
\begin{equation}
\begin{aligned}
 \CA_{2\dot2 A\dot B}\ =\ 0,\quad
 \CA_{1\dot2 A\dot B}\ =\ D_{2\dot 2 A\dot B}\Phi,\quad
 \CA_{2\dot1 A\dot B}\ =\ D_{2\dot 2 A\dot B}\Psi,\\
 \CA_{1\dot1 A\dot B}\ =\ D_{1\dot 2 A\dot B}\Psi
     +(D_{2\dot2 A\dot B}\Phi)\Psi+D_{2\dot1\dot A\dot B}\Phi+
       (D_{2\dot 2 A\dot B}\Psi)\Phi-D_{2\dot 2 A\dot B}\Sigma,\\
 \CA_{2\dot I}\ =\ 0,\qquad\CA_{1\dot I}\ =\ D_{2\dot I}\Phi,\qquad
 \CA_{I\dot 2}\ =\ 0,\qquad\CA_{I\dot1}\ =\ D_{I\dot 2}\Psi.
\end{aligned}
\end{equation}
Recall that for $A=1$, $\dot B=\dot1$ the
field $\CA_{2\dot2 A\dot B}$ is identified with the component 
$\CA_{2\dot2}$ of the Yang-Mills potential
$\CA_{\a\bd}:=\CA_{\a\bd 1\dot1}$. Since $\CA_{2\dot2}=0$, 
the terminology ``light-cone gauge"
becomes transparent from the viewpoint of $\CN=3$ SYM theory
(see also \eqref{eq:rcx2}, where ${\rm i}x^{2\dot2}=x^0-x^3$).

Finally, we stress that \eqref{eq:LCeq1} and \eqref{eq:LCeq2} 
contain the
equations
\begin{equation}\label{eq:GP2}
\begin{aligned}
D_{i1}D_{j2}\Phi+D_{j1}D_{i2}\Phi+
\{D_{i2}\Phi,D_{j2}\Phi\}\ &=\ 0,\\
D^i_{\dot1}D^j_{\dot2}\Psi+D^j_{\dot1}D^i_{\dot2}\Psi+
         \{D^i_{\dot2}\Psi,D^j_{\dot2}\Psi\}\ &=\ 0,\\ 
 D^i_{\dot2}\Sigma-D^i_{\dot1}\Phi-D^i_{\dot2}(\Psi\Phi)\ =\ 0,\qquad 
 D_{i2}\Sigma-D_{i1}\Psi-D_{i2}(\Phi\Psi)\ &=\ 0,\\
 D^i_{\dot2}\Phi\ =\ 0,\qquad D_{i2}\Psi\ &=\ 0          
\end{aligned}
\end{equation}
as a subset. These equations are equivalent to the field equations 
of
$\CN=3$ SYM theory in light-cone gauge. This is easily shown by
substituting an expansion of the form \eqref{eq:exp1}
into the linear system \eqref{eq:LSSYM} of $\CN=3$ SYM theory
as well as into the corresponding constraint equations \eqref{eq:CE}.

\section{Hidden symmetries of $\CN=3$ SYM theory} 

In this section, we consider the SYM hierarchy written
in light-cone gauge 
as an infinite set of equations on the prepotentials 
$\{\Phi,\Psi,\Sigma\}$
and discuss an algebra of hidden symmetries of the $\CN=3$ SYM
equations realized as point symmetries of the SYM hierarchy
via derivatives with respect to extra bosonic and fermionic
``times". In particular, we reinterpret the results derived in the
previous sections and in addition show how a ``double" affinization
of the algebra of supertranslations on Minkowski superspace can be
realized as a hidden symmetry algebra of $\CN=3$ SYM theory.

\paragraph{Linearized $\CN=3$ SYM equations.}
Let us consider the linearized form of 
\eqref{eq:GP2},
\begin{subequations}\label{eq:linLCG}
\begin{eqnarray}
 D_{(i1}D_{j)2}\d\Phi+\{D_{(i2}\Phi,D_{j)2}\d\Phi\}&=&0,\\
 D^{(i}_{\dot1}D^{j)}_{\dot2}\d\Psi+
\{D^{(i}_{\dot2}\Psi,D^{j)}_{\dot2}\d\Psi\}&=&0,\\     
 D^i_{\dot2}\d\Sigma-D^i_{\dot1}\d\Phi-
D^i_{\dot2}((\d\Psi)\Phi+\Psi\d\Phi) &=& 0,\\
D_{i2}\d\Sigma-D_{i1}\d\Psi-D_{i2}((\d\Phi)\Psi+\Phi\d\Psi)&=&0,\\
D^i_{\dot2}\d\Phi\ =\ 0,\qquad D_{i2}\d\Psi&=& 0.
\end{eqnarray}
\end{subequations}
Solutions $\{\d\Phi,\d\Psi,\d\Sigma\}$ to these equations 
for given $\{\Phi,\Psi,\Sigma\}$
satisfying \eqref{eq:GP2} are called (infinitesimal)
symmetries of the $\CN=3$ SYM
equations written in light-cone gauge. From the geometric 
point of view, solutions
to \eqref{eq:linLCG} are vector fields on the solution 
space of the
$\CN=3$ SYM equations. Besides natural local 
symmetries (e.g. generated by
superconformal and gauge transformations), the $\CN=3$ 
SYM equations possess infinitely
many nonlocal hidden symmetries, as we 
will show momentarily. Such symmetry
transformations generate new solutions from old ones. 
Here, we describe an
important subclass of such infinitesimal
symmetry transformations 
associated with an affinization
of supertranslations on $\IC^{4|12}$. These transformations 
generate the higher flows
in the SYM hierarchy introduced earlier in this paper.

\paragraph{Infinitesimal symmetries.}
Recall that the field equations of $\CN=3$ SYM theory 
\eqref{eq:GP2}
form a subset of the equations \eqref{eq:LCeq1}, 
\eqref{eq:LCeq2} of
the SYM hierarchy.\footnote{From now on, we assume
that all indices in 
\eqref{eq:LCeq1}, \eqref{eq:LCeq2} (besides R-symmetry 
indices $i,j$) run
from $1$ to $\infty$, i.e. they take values in $\IN$. 
In principle,
 one can also
allow them to take values in $\IZ$.} So, the prepotentials 
$\{\Phi,\Psi,\Sigma\}$
in \eqref{eq:GP2}, which
we symbolically denote by $\Theta$, depend not only on 
coordinates of 
Minkowski superspace but also on an infinite number of 
additional moduli
$x^{\a\,\dot b+\dot2}$, $x^{a+2\,\bd}$, $x^{a+2\,\dot b+\dot 2}$, 
$t^{a\dot B}$, 
$s^{A\dot b}$, $\eta^{\dot a+\dot2}_i$ and $\te^{i a+2}$.
 Therefore, upon
defining
\begin{subequations}\label{eq:sym1}
\begin{eqnarray}
\d^{(x)}_{A\dot B}\Theta\ :=\ D_{A\dot B\dot1}\Theta\ =\ 
        \der{x^{A\dot B}}\Theta, \label{eq:sym1a}\kern2.9cm\\
    \d^{(t)}_{a\dot B}\Theta\ :=\ D_{a\dot B\dot1}
                 \Theta\ =\   \der{t^{a\dot B}}\Theta,
    \qquad \d^{(s)}_{A\dot b}\Theta\ :=\ D_{1A\dot b}\Theta\ =\
        \der{s^{A\dot b}}\Theta,\label{eq:sym1b}\\
     \d_{iA}\Theta\ :=\ \left(D_{iA}-
       2\eta^{\dot B}_i\der{x^{A\dot B}}\right)\Theta\ 
=:\ Q_{iA}\Theta,
       \label{eq:sym1c}\kern1.9cm\\       
  \d^i_{\dot A}\Theta\ :=\ \left(D^i_{\dot A}-
      2\te^{iB}\der{x^{B\dot A}}\right)\Theta\ =:\ 
Q^i_{\dot A}\Theta,
     \label{eq:sym1d}\kern2cm
\end{eqnarray}
\end{subequations}
we obtain infinitesimal symmetries of the 
equations 
\eqref{eq:GP2}
since all vector fields appearing in \eqref{eq:sym1} 
(anti)commute with 
$D_{i\a}$ and $D^i_\ad$. The fermionic vector fields 
$Q_{iA}$ and
$Q^i_{\dot A}$ introduced above obey
\begin{equation}
 \big\{Q_{iA},Q^j_{\dot B}\big\}\ =\ -2\d^j_i\der{x^{A\dot B}}
\end{equation}
and in addition they also anticommute with $D_{iA}$ and 
$D^i_{\dot A}$.
Thus, the infinitesimal transformations $\Theta\mapsto\d\Theta$
defined in \eqref{eq:sym1} as derivatives with respect to moduli
parametrizing higher flows of the SYM hierarchy give solutions 
to the
linearized $\CN=3$ SYM equations in light-cone gauge 
\eqref{eq:linLCG},
i.e. they define an infinite set of hidden (infinitesimal)
symmetries.  

\paragraph{Symmetry equations as subset of the SYM hierarchy.}
Recall that the derivatives in \eqref{eq:sym1} generate 
bosonic and
fermionic flows on the solution space of $\CN=3$ SYM theory. 
This
hints that equations \eqref{eq:linLCG} on infinitesimal
symmetries defined by
\eqref{eq:sym1} may follow from a subset of the equations
of the SYM hierarchy. We will show that this is 
indeed
the case. Moreover, the remaining equations of the hierarchy 
describe
the dependence of the symmetries $\d\Theta$ 
on the additional moduli. This in fact is in the same spirit 
as for
self-dual (S)YM hierarchies (see e.g. 
\cite{MasonRF,IvanovaZT,Wolf:2004hp}). We exemplify all this by
focussing on
\begin{equation}\label{eq:sym2}
 \d_{2\dot\Upsilon}\Theta\ :=\ D_{2\dot\Upsilon}\Theta,
\end{equation}
which, due to \eqref{eq:TVF1} and \eqref{eq:EVF}, are linear 
combinations
of the symmetries \eqref{eq:sym1a}, \eqref{eq:sym1b}.

Let us substitute \eqref{eq:sym2} into the equations 
\eqref{eq:LCeq1b}, \eqref{eq:LCeq1c} and \eqref{eq:LCeq2b}
with $A=1$. Remember also that
$D_{i1A}\equiv D_{iA}$, $D_{i2A}\equiv D_{iA+1}$, 
$D^i_{\dot1\dot A}\equiv D^i_{\dot A}$ and 
$D^i_{\dot2\dot A}\equiv D^i_{\dot A+\dot1}$. Then we have the 
following
equations on the symmetries \eqref{eq:sym2}:
\begin{subequations}\label{eq:sym3}
\begin{eqnarray}
  D_{j2}\d_{1\dot\Upsilon}\Phi-D_{j1}\d_{2\dot\Upsilon}\Phi+
 [\d_{2\dot\Upsilon}\Phi,D_{j2}\Phi]&=&0,\label{eq:sym3c}\\
 \d_{2\dot\Upsilon}\Sigma-\d_{1\dot\Upsilon}\Psi-
  (\d_{2\dot\Upsilon}\Phi)\Psi&=&0,\label{eq:sym3b}\\
  \d_{2\dot\Upsilon}\Psi&=&0.\label{eq:sym3a}
\end{eqnarray}
\end{subequations}
Here,
\begin{equation}\label{eq:sym4}
 \d_{1\dot\Upsilon}\Theta\ :=\ D_{1\dot\Upsilon}\Theta
\end{equation}
are also linear combinations of the symmetries \eqref{eq:sym1a},
\eqref{eq:sym1b}. Applying 
$D_{i2}$ and $D^i_{\dot2}$ to \eqref{eq:sym3b},
$D_{i2}$ to \eqref{eq:sym3c}, and symmetrizing the
R-symmetry indices $i$ and $j$, we find
\begin{subequations}\label{eq:sym5}
\begin{eqnarray}
 D_{(i1}D_{j)2}\d_{2\dot\Upsilon}\Phi+\{D_{(i2}\Phi,
D_{j)2}\d_{2\dot\Upsilon}\Phi\}&=&0,\\
 D^{(i}_{\dot1}D^{j)}_{\dot2}\d_{2\dot\Upsilon}\Psi+
\{D^{(i}_{\dot2}\Psi,D^{j)}_{\dot2}
\d_{2\dot\Upsilon}\Psi\}&=&0,\\
D_{i2}\d_{2\dot\Upsilon}\Sigma-D_{i1}\d_{2\dot\Upsilon}
\Psi-D_{i2}((\d_{2\dot\Upsilon}\Phi)\Psi+
\Phi\d_{2\dot\Upsilon}\Psi)&=&0,\\
 D^i_{\dot2}\d_{2\dot\Upsilon}\Sigma-D^i_{\dot1}
\d_{2\dot\Upsilon}\Phi-D^i_{\dot2}((\d_{2\dot\Upsilon}
\Psi)\Phi+\Psi\d_{2\dot\Upsilon}\Phi) &=& 0,\\
D_{i2}(\d_{2\dot\Upsilon}\Psi)\ =\ 0,\qquad 
D^i_{\dot2}(\d_{2\dot\Upsilon}\Phi) &=& 0,
\end{eqnarray}
\end{subequations}
where we used \eqref{eq:LCeq1}, \eqref{eq:LCeq2} and 
\eqref{eq:sym3a}. Obviously, \eqref{eq:sym5} is a
special case of \eqref{eq:linLCG} for the symmetries
\eqref{eq:sym2}. The remaining equations in 
\eqref{eq:LCeq1}, \eqref{eq:LCeq2} on the symmetries
\eqref{eq:sym2}, e.g.
\begin{equation}
 D_{1\dot\Upsilon}\d_{2\dot\Lambda}\Phi-D_{1\dot\Lambda}
 \d_{2\dot\Upsilon}\Phi
 +[D_{2\dot\Lambda}\Phi,\d_{2\dot\Upsilon}\Phi]\ =\ 0,
\end{equation}
describe their dependence on the moduli along the higher
flows. Similarly, one can show for all symmetries
\eqref{eq:sym1} (and their linear combinations) that 
the linearized $\CN=3$ SYM equations  \eqref{eq:linLCG}
can be obtained from a subset of the SYM hierarchy
\eqref{eq:LCeq1}, \eqref{eq:LCeq2}. Note that the
infinitesimal transformations \eqref{eq:sym1} were
defined to be compatible with \eqref{eq:RCTheta},
which are the reality conditions for the flows.

\paragraph{Symmetry algebra.}
As an immediate consequence of the transformations 
\eqref{eq:sym1}, we find
\begin{equation}
 \big\{\d_{iA},\d^j_{\dot B}\big\}\ =\ -2\d^j_i
   \d^{(x)}_{A\dot B}
\end{equation}
upon acting on either $\Phi$, $\Psi$ or $\Sigma$.
These relations are the only nonvanishing (anti)commutators
of two successive infinitesimal
symmetry transformations. We have thus
obtained an infinite-dimensional (graded) algebra of
hidden symmetries of the $\CN=3$ SYM equations.

To understand the nature of this algebra, let us consider
the action of the vector fields entering in the
definition of the transformations \eqref{eq:sym1} on the
transition functions $\tilde f=\{\tilde f_{pq}\}$ of
the holomorphic vector bundle $\tilde\CE\to\CF^{M+2|N}$
for finite $n$ and eventually put $n\to\infty$.
A short calculation reveals that on the intersection
$\hCW_1\cap\hCW_p$ we have
\begin{equation}\label{eq:receq}
\begin{aligned}
 \mu^a\lambda^{\dot b}\der{x^{1\dot1}}\ &=\
 \der{x^{a+1\,\dot b+\dot1}}+\der{t^{a\,\dot b+\dot1}}+
 \der{s^{a+1\,\dot b}},\\
 \mu^a\lambda^{\dot b}\der{x^{1\dot2}}\ &=\
 \der{x^{a+1\,\dot b+\dot2}}+\der{t^{a\,\dot b+\dot2}},\\
 \mu^a\lambda^{\dot b}\der{x^{2\dot1}}\ &=\
 \der{x^{a+2\,\dot b+\dot1}}+\der{s^{a+2\,\dot b}},\\
 \mu^a\lambda^{\dot b}\der{x^{2\dot2}}\ &=\
 \der{x^{a+2\,\dot b+\dot2}},\\
 \mu^a Q_{i1}\ &=\ Q_{ia+1}-2\eta^{\dot B}_i\der{t^{a\dot B}},
 \qquad \mu^a Q_{i2}\ =\ Q_{ia+2},\\
 \lambda^{\dot a}Q^i_{\dot1}\ &=\ Q^i_{\dot a+\dot1}-2\te^{iB}
 \der{s^{B\dot a}},\qquad\lambda^{\dot a}Q^i_{\dot2}\ =\ 
 Q^i_{\dot a+\dot2},
\end{aligned}
\end{equation}
where the above relations are understood upon action on
$\tilde f_{1p}$. Similar expressions can be derived for
the other patches. Therefore, we see that the vector fields
on the right-hand side of \eqref{eq:receq} are 
recursively generated from the generators of supertranslations
on Minkowski superspace. Recall that for the real case, we have
the relations \eqref{eq:rcx1}--\eqref{eq:rcx2} and $\mu=\bar\lambda$.

We see that the obtained algebra is related to the 
algebra $T^{4|12}[\lambda,\mu]$ with generators
\begin{equation}\label{eq:gen2}
  \mu^a\lambda^{\dot b}P_{\a\bd},\qquad
       \mu^a Q_{i\a}\qquad{\rm and}\qquad
     \lambda^{\dot b} Q^i_\bd
\end{equation}
for $a,\dot b\in\IN$.
Here, $P_{\a\bd}:=\partial/\partial x^{\a\bd}$ and 
$T^{4|12}$ denotes the supertranslation algebra generated
by $P_{\a\bd}$, $Q_{i\a}$ and $Q^i_\bd$. As shown above,
the algebra $T^{4|12}[\lambda,\mu]$ is 
represented via \eqref{eq:receq} in terms of vector fields
on the solution space of the $\CN=3$ SYM equations. Altogether,
we have thus obtained an affinization of the algebra of
supertranslation $T^{4|12}$ on complexified Minkowski
superspace. Moreover, in view of 
the self-dual bosonic case \cite{MasonRF},
it seems conceivable that by considering all four 
patches which cover  
generalized superambitwistor space and by extending 
the range of all indices
$a,\dot a,\ldots$ in the SYM hierarchy to all integers $\IZ$,
 the above algebra can be extended to the
symmetry algebra $T^{4|12}[\lambda,\lambda^{-1},\mu,\mu^{-1}]$
which is generated by the same set of generators \eqref{eq:gen2}
but for  $a,\dot b\in\IZ$. Finally, we note that
in the real case, the graded double loop
algebra $T^{4|12}[\lambda,\lambda^{-1},\mu,\mu^{-1}]$ 
becomes $T^{4|12}[\lambda,\lambda^{-1},\bar\lambda,\bar\lambda^{-1}]$
with the identity $\mu^a Q_{i\a}=(\lambda^{\dot a}Q^i_\ad)^\dagger$,
etc. for $\bar{\dot a}=a$.

\section{Summary and discussion}

The goal of this paper was to present a step towards an 
understanding of the appearance of nonlocal hidden symmetries
in $\CN=4$ SYM theory from first principles. Here,
we were interested in symmetries related to particular
space-time symmetries. For this, we
introduced a generalized superambitwistor space 
$\CL^{5|6}_{m,n}$ fibred over $\IC P^1\times\IC P^1_*$
together with the space $\IC^{4|12}\times\IC^{M-4|N-12}$
parametrizing subspaces $(\IC P^1\times\IC P^1_*)_\tx$ in
$\CL^{5|6}_{m,n}$ with $\tx\in\IC^{M|N}$. Then, by specializing
to the case $\CL^{5|6}_n:=\CL^{5|6}_{n,n}$, we discussed a
 Penrose-Ward type
transform relating holomorphic vector bundles over
$\CL^{5|6}_n$ and solutions to Yang-Mills-Higgs type
equations on $\IC^{M|N}=\IC^{3n^2+2n-1|6(n+1)}$ being termed 
the SYM($n$) equations which describe the SYM hierarchy 
truncated up to level $n$. Note that $\CL^{5|6}_1$ coincides
with the well-known superambitwistor space $\CL^{5|6}$ and
the SYM(1) equations are equivalent to the equations of 
motion of $\CN=3$ SYM theory. The truncated SYM hierarchy
turns into the full SYM hierarchy after taking the asymptotic
limit $n\to\infty$. The field equations of $\CN=3$ SYM theory
are embedded into this infinite system of partial differential
equations as well as into the SYM($n$) equations for finite
$n>1$. Hence, a given solution to the $\CN=3$ SYM equations
can be embedded into an infinite-parameter family of
solutions to the whole hierarchy.\footnote{This is a generic
situation. However, for some concrete solutions, there could
be obstructions of the same nature as for the SDYM
hierarchy discussed in \cite{MasonRF}.} The dependence of the
fields on the extra parameters can be recovered by solving
the equations of the SYM hierarchy.

Furthermore, we reinterpreted the equations of the SYM 
hierarchy in the context of hidden symmetries of $\CN=3$
SYM theory. In particular, we rewrote the constraint equations
of the latter and those of the hierarchy in light-cone gauge.
In this gauge, parametrized by three prepotentials $\Phi$, $\Psi$
and $\Sigma$, it became transparent that some equations of the
SYM hierarchy are equations on hidden 
infinitesimal symmetries (i.e.
solutions to the linearized field equations) of $\CN=3$ SYM
theory. We have shown that these nonlocal symmetries are related
to a graded algebra $T^{4|12}[\lambda,\mu]$ generated from
the algebra $T^{4|12}$ of supertranslations on Minkowski 
superspace. In this respect, we emphasize
again that the $\CN=3$ and $\CN=4$ SYM theories
are physically equivalent theories.
Thus, the twistor description gives a direct way 
of obtaining a particular set of infinitely many hidden
symmetries in $\CN=4$ SYM theory.

Recall that the classical Green-Schwarz superstring on the
curved background AdS$_5\times S^5$ possesses an infinite set
of hidden symmetries and nonlocal conserved charges 
\cite{Bena:2003wd}
which are similar to those in two-dimensional models
(see e.g. \cite{LuscherRQ}). This is basically because of the
possibility of interpreting the Green-Schwarz superstring on
AdS$_5\times S^5$ as a particular sigma model, 
where the fields
take values in the coset space $PSU(2,2|4)/(SO(1,4)\times SO(5))$.
By virtue of the AdS/CFT correspondence \cite{Maldacena:1997re},
conserved nonlocal charges should also exist in $\CN=4$ SYM
theory (at least in the planar limit). In fact, within the
spin chain approach, the authors
of \cite{Dolan:2003uh} were able to derive an analogous set
of conserved charges in the superconformal Yang-Mills theory in the
weak coupling limit. It is quite reasonable to expect that these
Yangian symmetries and charges could be related to those
derived in the present work.

As was mentioned in section 1, in recent years essential 
progress in our understanding of quantum properties of
$\CN=4$ SYM theory has been made with the help of twistor
string theory \cite{Witten:2003nn}. Besides B-type topological
string theory on the supertwistor space $\CP^{3|4}$, Witten
also mentioned the possibility of formulating a twistor
string theory on the superambitwistor space\footnote{Both 
spaces $\CP^{3|4}$ and $\CL^{5|6}$
are Calabi-Yau supermanifolds, a property which is essential
for using them as target spaces for B-type topological string
theory.} 
$\CL^{5|6}$. Within such a formulation the mechanism of reproducing
perturbative $\CN=4$ SYM theory would be completely
different compared to the supertwistor space approach (i.e.
no D-instantons are needed), as already
at classical level, holomorphic Chern-Simons theory on 
superambitwistor
space yields all the interactions of 
$\CN=4$ SYM theory.
However, there are problems in formulating an action principle 
for
holomorphic Chern-Simons theory on this space caused by the 
difficulty in making sense of an appropriate integration measure. 
Recently, some progress in this direction has been made in 
\cite{Mason:2005kn} for Euclidean signature in four dimensions.

Note that generalized superambitwistor space $\CL^{5|6}_{m,n}$ is
a Calabi-Yau supermanifold for any values of $m$ and $n$. One can
therefore define a twistor string theory on $\CL^{5|6}_{m,n}$. It
is also an interesting task to see how the ideas presented in 
\cite{Mason:2005kn}
should be generalized in order to construct appropriate action
functionals for the truncated SYM($n$) hierarchies introduced in
our paper. The next obvious step is the generalization of the
twistor construction to all generators of the superconformal
group. Besides questions associated with symmetry transformations,
one in addition needs to write down the related conserved currents
and charges (not only as superfields but also in components) in
order to proceed further to quantum theory. It will then hopefully 
be possible to make contact with the quantum symmetry algebras
considered in \cite{Dolan:2003uh}. Moreover, it would also be
interesting to see whether the symmetries described in this paper
have an analog in the context of twistor string theory.

\bigbreak\bigskip\noindent{\bf Acknowledgements}
\vspace*{.3cm}

We would like to thank O. Lechtenfeld and R. Wimmer for 
useful discussions. This work was partially supported by
the Deutsche Forschungsgemeinschaft (DFG).

\end{document}